\documentclass[12pt,preprint]{aastex}
\def\msun{\rm M_{\sun}}

\def\av{${\rm A_V}$}

\begin{document}
\shortauthors{Hernandez et al.}
\shorttitle{ {\em Spitzer} in the $\sigma$ Orionis cluster}

\title{A {\em Spitzer Space Telescope} study of disks in the young $\sigma$ Orionis cluster}

\author{Jes\'{u}s Hern\'andez\altaffilmark{1,2}, L. Hartmann\altaffilmark{1}, T. Megeath\altaffilmark{3},
R. Gutermuth \altaffilmark{4}, J. Muzerolle\altaffilmark{5}, N. Calvet\altaffilmark{1},
A. K. Vivas\altaffilmark{2}, C. Brice\~{n}o\altaffilmark{2}, L. Allen\altaffilmark{4},
J. Stauffer\altaffilmark{6}, E. Young\altaffilmark{5} and G. Fazio\altaffilmark{4}}

\altaffiltext{1}{Department of Astronomy, University of Michigan, 830 Dennison Building, 500 Church Street, Ann Arbor, MI 48109, US}
\altaffiltext{2}{Centro de Investigaciones de Astronom\'{\i}a, Apdo. Postal 264, M\'{e}rida 5101-A, Venezuela.}

\altaffiltext{3}{Ritter Observatory, MS 113, Department of Physics and Astronomy, University of Toledo, OH 43606-3390.}

\altaffiltext{4}{Harvard-Smithsonian Center for Astrophysics, 60 Cambridge, MA 02138, US}

\altaffiltext{5}{Steward Observatory, University of Arizona, 933 North Cherry Avenue, Tucson, AZ 85721, US}

\altaffiltext{6}{{\em Spitzer} Science Center, Caltech M/S 220-6, 1200 East California Boulevard, Pasadena, CA 91125}

\email{hernandj@umich.edu}

\begin{abstract}
We report new {\em Spitzer Space Telescope} observations from the IRAC
and MIPS instruments of the young ($\sim$ 3 Myr) $\sigma$ Orionis cluster.
The populous nature of this cluster makes it a good target for statistically-significant
studies of disk emission as a function of mass.
We identify 336 stars as members of the cluster using optical and near-infrared
color magnitude diagrams.
Using the spectral energy distribution (SED) slopes in the IRAC spectral range,
we place objects in several classes:
non-excess stars, stars with optically thick disks (like classical T Tauri stars),
class I (protostellar) candidates, and stars with ``evolved disks''; the last
exhibit smaller IRAC excesses than optically thick disk systems.
In general, this classification agrees with the location expected in IRAC-MIPS
color-color diagrams for these objects.
We find that the evolved disk systems are mostly a combination of objects with
optically thick but non-flared disks, suggesting grain growth and/or settling,
and transition disks, systems in which the inner disk is partially or fully cleared
of small dust. 
In all, we identify 7 transition disk candidates and 3 possible
debris disk systems.
There appears to be a spatial extension of infrared excess sources 
to the north-east, which may be associated with
the young ($<1$ Myr) embedded cluster NGC 2024.
As in other young stellar populations,
the fraction of disks depends on the stellar mass, ranging from $\sim$10\% for
stars in the Herbig Ae/Be mass range ($>$2 $\msun$) to $\sim$35\% in the
T Tauri mass range (1-0.1 $\msun$). 
We find that the disk fraction does not decrease significantly
toward the brown dwarf candidates ($<$0.1 $\msun$).
The IRAC infrared excesses found in stellar clusters and associations 
with and without central high mass stars are similar, suggesting that 
external photoevaporation is not very important in many clusters.
Finally, we find no correlation between the X-ray luminosity and the disk
infrared excess, suggesting that the X-rays are not strongly affected by
disk accretion.

\end{abstract}


\section{Introduction}
\label{sec:int}

Observations have shown and theory predicts 
that disks around young stars are an inescapable consequence of star formation,
given angular momentum conservation during protostellar cloud core collapse
\citep{hartmann05a,ward05,andre00}. As the resulting protoplanetary
disks age, their excesses at near-infrared (NIR) and mid-infrared wavelengths
decrease.  Previous studies have indicated that
the timescale for this evolution in disk emission is strongly dependent on
the stellar mass \citep{lada95,muzerolle03,calvet04,aurora05}
ranging from 5-7 Myr for objects in the mass range of the T Tauri stars 
\citep[ TTS, with spectral type K5 or later; ][]{haisch01, hartmann05b,hillenbrand06}
to $<$3 Myr for objects in the mass range of the Herbig Ae/Be (HAeBe)
stars \citep[spectral type F5 or earlier;][]{hernandez05}.

The observed decreases in infrared emission can result from:
grain growth to sizes much larger than the wavelength of observation;
dust settling to the disk midplane, which reduces the ``flaring'' of
the disk and thus the amount of energy radiated \citep{kh87, dullemond05, dalessio06};
clearing of small dust particles by large bodies in the disk; or some
combination of these processes.
Since grain growth and settling to the mid-plane occur fastest in the 
(warmer) inner disk \citep{weidenschilling97, dullemond04}, the 
disk emission is expected to decline faster at short wavelengths than
at long wavelengths, and there is observational evidence for this
\citep{aurora06,hartmann05b,lada06}.

Of particular interest are "transition disk" objects,
which have an inner, optically thin disk region,
possibly produced by planet clearing, combined with an outer, optically
thick disk \citep{calvet02,calvet05,dalessio05a, megeath05a}.  Prior to
the complete or nearly-complete clearing seen in transition disks, 
one would expect to observe optically-thick but reduced infrared emission
from the inner disk, for which there is some evidence 
\citep[e.g., ][]{aurora06, lada06}. However, additional large samples
of stars are desirable to make the result more statistically
significant, as stars of the same mass and age show a wide
range of disk emission properties \citep{furlan06}.

The $\sigma$ Orionis cluster is of particular interest
because it is reasonably near and relatively populous, making
possible statistically-significant studies of disk properties
as a function of stellar mass.  Moreover, $\sigma$ Orionis
has an age $\sim 3$~Myr at which one might expect the beginnings of disk
evolution to become evident.  It is part of the 
Orion OB1b sub-association which has an age of 1.7-7 Myr\citep{warren78,brown94,brown98,
briceno05}.  Low-mass members of the $\sigma$ Orionis cluster were first reported
by \citet{walter97}, who found over 80 X-ray sources
and spectroscopically identified more than 100 low mass, pre-main
sequence (PMS) members lying within 1 degree from the star $\sigma$ Ori.
The distance calculated by Hipparcos for this star (352$^{+166}_{-85}$) agrees,
within the uncertainties of Hipparcos, with the Hipparcos distance calculated
for the overall population of the stellar subassociation OB1b
\citep[439$\pm$33;][]{brown98}, which is statistically more reliable. 
The estimated age of the cluster is 2-4 Myr
\citep{zapatero02,oliveira02,sherry04,franciosini06}. Since
the $\sigma$ Orionis cluster is relatively near and the reddening toward the
center of the cluster is low \citep[E(B-V)=0.05 mag; ][]{brown94,bejar99},
this stellar cluster is an excellent laboratory to study
young stars in a entire range of mass, from  the massive and multiple
O9.5 type star $\sigma$ Ori to the lowest mass objects, like brown dwarfs
and free-floating planets \citep[e.g.,][]{barrado01,barrado03,bejar99,
bejar01,bejar04,burningham05,
caballero05, kenyon05,oliveira06, scholz04,sherry04,zapatero02,walter98}.

The unprecedented sensitivity and spatial resolution provided by
the {\em Spitzer Space Telescope} in the near- and mid-infrared
windows are powerful tools to expand significantly our understanding
of star and planet formation processes.
In this contribution, we analyze the near- and mid-infrared  properties of stars in
the $\sigma$ Orionis cluster, ranging in mass  from HAeBe stars to the substellar limit.

This paper is organized as follows. In \S \ref{sec:obs} we present the observational
data.  In \S \ref{sec:selmem} we describe the selection of cluster
members. We analyze the observations and describe the results on
\S \ref{sec:res}, and give our conclusions in \S \ref{sec:conc}.

\section{Observations}
\label{sec:obs}
\subsection{Infrared photometry}
\label{sec2:ir}
We have observed a field of $\sim$1730 arcmin$^2$ on the $\sigma$ Orionis cluster using 
the four channels (3.6, 4.5, 5.8, and 8.0 \micron) of the InfraRed Array Camera 
\citep[IRAC, ][]{fazio04}; 90\% of this field was also observed using 
the 24 {\micron} band of the Multiband Imaging Spectrometer for {\em Spitzer} \citep[MIPS, ][]{rieke04}.

The IRAC observations presented here were taken on October 9, 2004.
The field of view was covered by a $9 \times 10$ position mosaic
(20'' overlap), with three dithered exposures at each position.
Images were obtained in High Dynamic Range (HDR) mode, whereby a
short integration (1 second) is immediately followed by a long
integration (26.8 seconds). Standard Basic Calibrated Data (BCD)
products from version S13.2 of the {\em Spitzer} Science Center's IRAC
pipeline were used to make the final mosaics. Post-BCD data
treatment was performed using custom IDL software (Gutermuth et al.
2004, Gutermuth et al. 2006) that includes modules for detection and
correction of bright source artifacts, detection and removal
of cosmic ray hits, construction of the long and short exposure
HDR mosaics, and the merger of those mosaics to yield the final
science images.

MIPS observations were made using the medium scan mode with half-array cross scan overlap, 
resulting in a total effective exposure time of 80 seconds.  The 24 {\micron} images 
were processed using the MIPS instrument team Data Analysis Tool (DAT), which calibrates
the data and applies a distortion correction to each individual exposure
before combining into a final mosaic \citep{gordon05}.

\begin{figure}
\epsscale{1.0}
\plotone{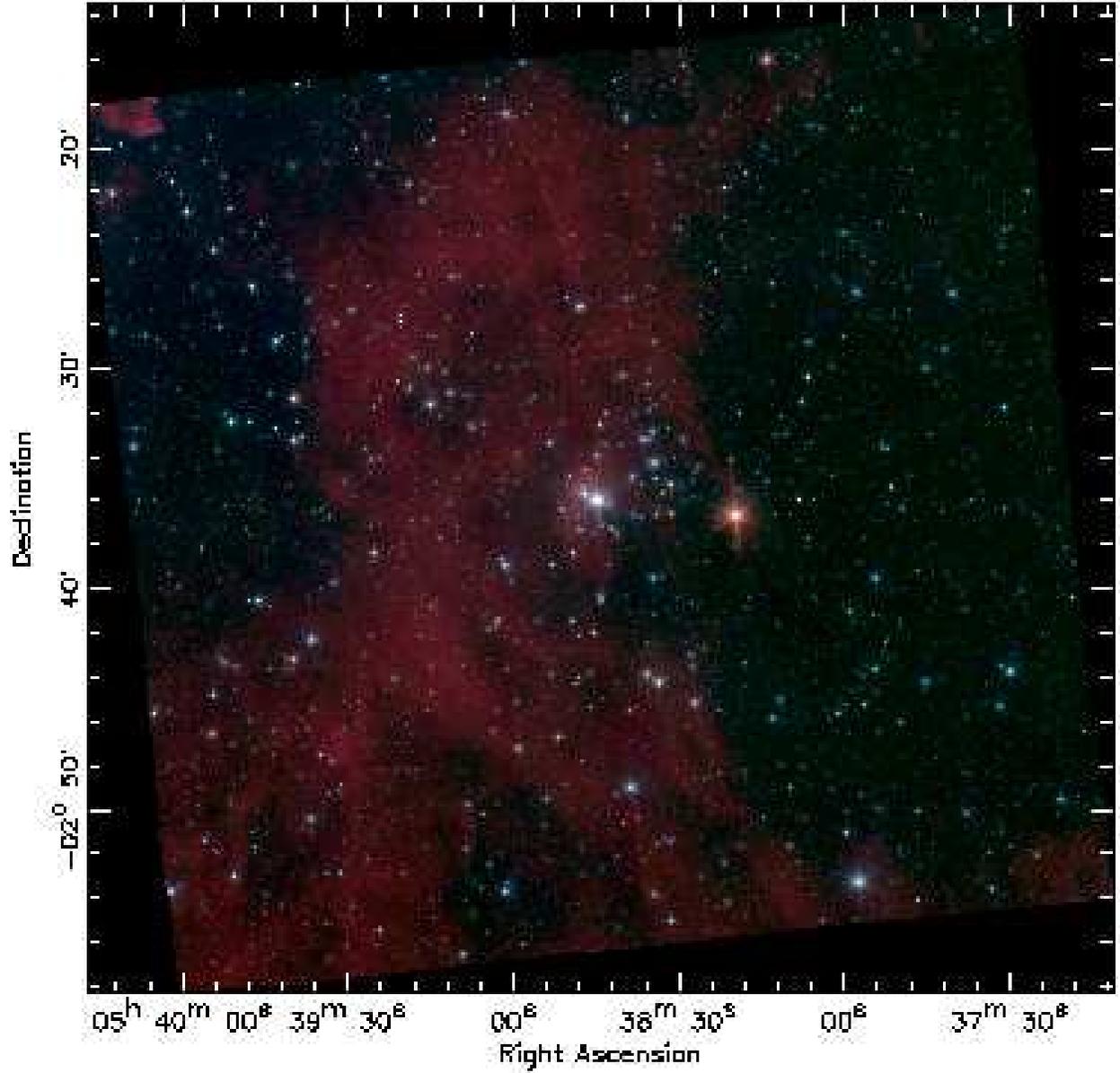}
\caption{False-color image of the $\sigma$ Orionis cluster. It is a 3-color composite of IRAC images, 
3.6 (blue), 4.5 (green), and 8.0 (red) \micron. 
The field is centering at the star $\sigma$ Ori.} 
\label{fig:field}
\end{figure}

\begin{figure}
\epsscale{0.8}
\plotone{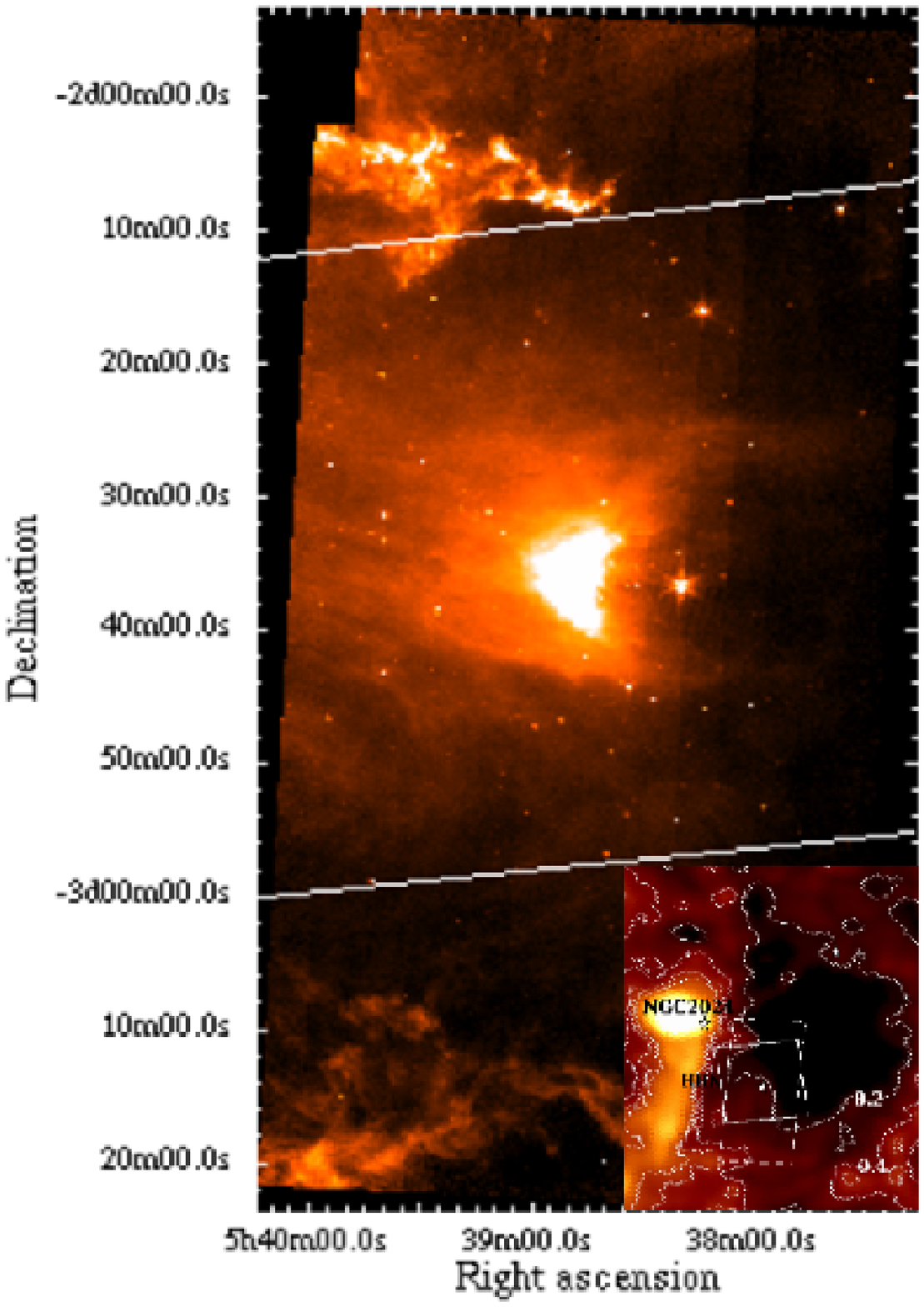}
\caption{MIPS 24 {\micron} image of the $\sigma$ Orionis cluster. 
The solid diagonal lines indicate the IRAC field.
The lower-right panel shows a map of dust infrared emission \citep{schlegel98}  
illustrating the location of the cluster in the OB1b sub-association 
\citep[ring-like structure; ][]{briceno05}; the isocontours indicate 
levels of galactic extinction (\av=0.2, 0.4, 0.6 \& 0.8 mag). The IRAC 
(small box) and MIPS (large box) fields are shown in the inset. The
positions of the  young stellar cluster ($<$1 Myr)
NGC 2024 and the Horsehead nebula are also indicated in this figure}
\label{fig:field2}
\end{figure}

Figures \ref{fig:field} and \ref{fig:field2} show the IRAC color and the MIPS 24 {\micron} images 
of the $\sigma$ Orionis cluster.
The lower-right panel in Figure \ref{fig:field2}
shows a map of dust infrared emission \citep{schlegel98} 
illustrating the location of the cluster in the OB1b sub-association, which can 
be traced by the ringlike structure centering at RA$\sim$84 and DEC$\sim$-2 
\citep{briceno05, hernandez06}.  In this panel, the boxes show
the positions of the IRAC (small box) and MIPS (large box) scan regions.
The isocontours are estimators of galactic extinction (for \av=0.2, 0.4, 0.6 \& 0.8 mag)
from the 100 {\micron} map of dust emission \citep[see, ][]{schlegel98}.
The galactic extinction and the background at 8 and 24 {\micron} 
is quite low compared with other regions with similar ages, which improves the precision 
and sensitivity of the photometry.

Point source detections were carried out individually on 
each IRAC band using PhotVis (version 1.09),
an IDL GUI-based photometry visualization tool developed by R. Gutermuth using the DAOPHOT modules 
ported to IDL as part of the IDL Astronomy User Library \citep{landsman93}. More 
than 10,000 sources were detected in at least one {\em Spitzer} band. 
We extracted the photometry of these objects using the {\it apphot} package in IRAF, 
with an aperture radius of 3\arcsec.7 and a background annulus from 3.7 to 8\arcsec.6.  
We adopted zero-point magnitudes for the standard aperture radius (12\arcsec) and 
background annulus (12-22\arcsec.4) of 19.665, 18.928, 16.847 and 17.391 in the 
[3.6], [4.5], [5.8] and [8.0] bands, respectively. 
Aperture corrections were made using the values described in IRAC Data Handbook \citep{reach06}. 
A preliminary list of 1682 objects (hereafter {\it Sample 1}) was created by selecting those 
objects having photometric measurements in all IRAC bands with errors less 
than 0.15 magnitudes.
We obtained point source photometry at 24 {\micron} with IRAF/{\it daophot}
point spread function fitting, using an aperture size of about 5\arcsec.7 and
an aperture correction factor of 1.73 derived from the STinyTim PSF model.
The absolute flux calibration uncertainty should be 4\% \citep{engelbracht06}.
Because of the contamination by the bright star $\sigma$ Ori 
in the IRAC and MIPS images, stars located within 30{\arcsec} of this O star were not included in 
the preliminary list.

\subsection{Optical photometry}
\label{sec2:opt}
Optical magnitudes were obtained from the CIDA Equatorial Variability survey
which is being carried out using the QUEST I camera \citep{baltay02} 
installed at the Jurgen Stock Telescope (a  1-m Schmidt telescope) at the Venezuelan 
National Observatory of Llano del Hato. The camera, an array of 4x4 CCDs, 
is designed to work in driftscan mode, which is a very efficient way to 
survey large areas of the sky. Each scan of the sky was reduced and calibrated 
with the standard QUEST software and the method described in  \citet{vivas04}
for scans centered at $\delta$=-1. We used data from multiple scans centered 
at $\delta=-3\degr$. The scans have a width of 2.2$^{\circ}$  in $\delta$. 
The range in right ascension shown in Figure 1 is covered completely by 
these optical observations. However, there is no data in a small region 
from -2.52$^{\circ}$~$<\delta <$~-2.44$^{\circ}$,
corresponding to the gap between CCDs in the 
QUEST I camera. Most  objects with an optical counterpart have more 
than 10 measurements, so we can identify the variables using differential 
photometry \citep{vivas04}.

\section{Membership selection}
\label{sec:selmem}

We do not have spectroscopic confirmation for most of the probable members
in the $\sigma$ Orionis cluster.  We must use photometric criteria to make a
robust selection of members, as described below.

\subsection{Rejecting extragalactic objects}
\label{sec:rejgalaxy}

\begin{figure}
\epsscale{0.9}
\plotone{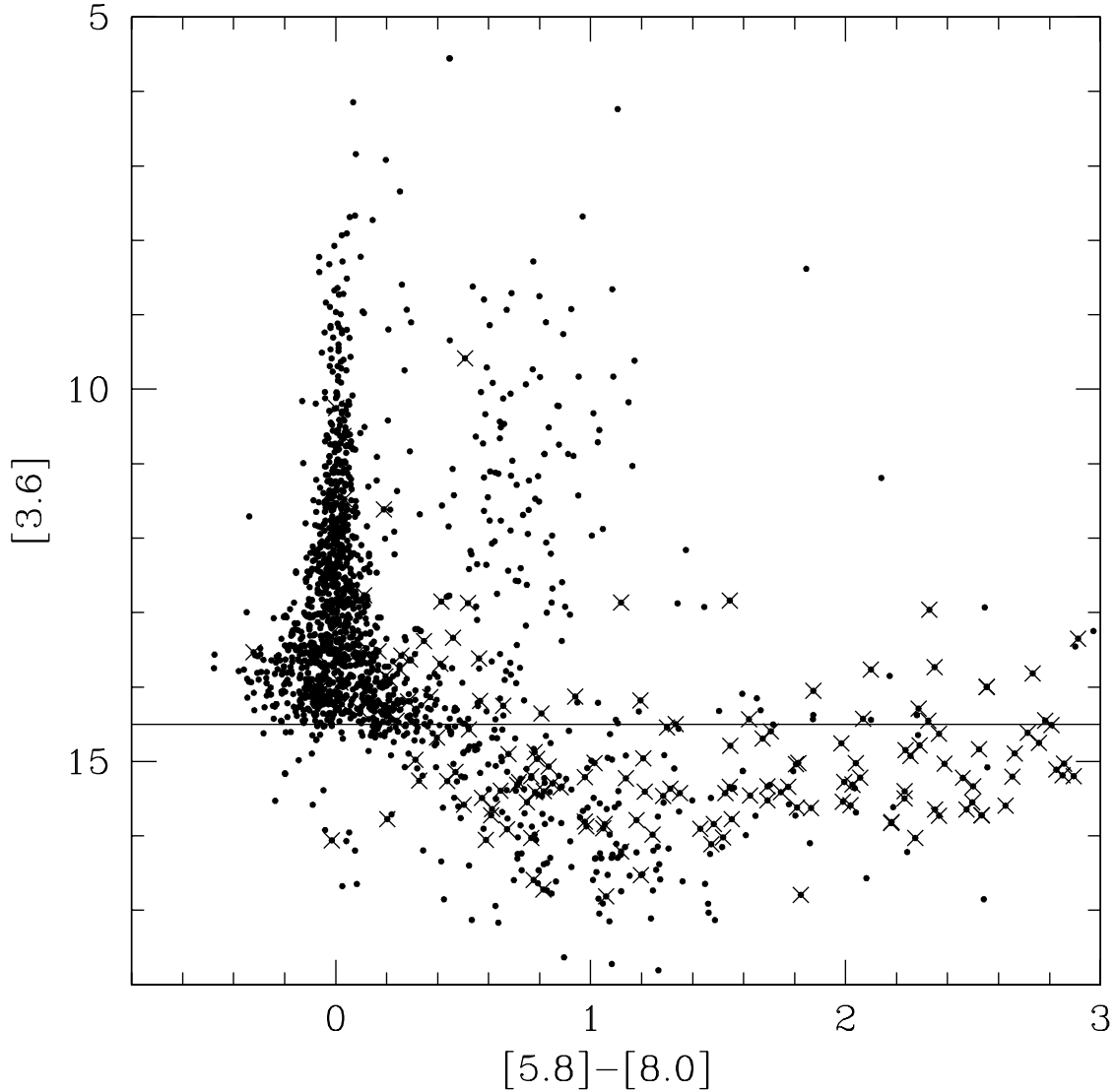}
\caption{[5.8]-[8.0] versus [3.6] color-magnitude diagram.
Objects with large FWHM in the radial profile measured in all IRAC bands
are likely to be extended sources (crosses). The density of objects with photospheric
colors ([5.8]-[8.0]$\sim$0)
decreases drastically for sources with [3.6]$>$14.5, most objects below this limit
(solid line) are extended sources. We selected objects above the solid line, where the
contamination from extragalactic sources is expected to be less than 50\% \citep{fazio04}.}
\label{fig:agn1}
\end{figure}

Figure \ref{fig:agn1} shows the [3.6] versus [5.8]-[8.0] color-magnitude diagram
for {\it Sample 1} (\S \ref{sec2:ir}).  The bulk of the sample has  [5.8]-[8.0]$\sim$0 and [3.6]$<$14.5;
below this limit the number of objects with photospheric colors ([5.8]-[8.0]$\sim$0)
decreases drastically,
and the contamination from extragalactic sources is expected to be more than 50\%
\citep{fazio04}.
We identify extended sources (crosses) by computing the radial profile for each object
using the {\it radprof} task in IRAF, and designate as extended sources
those with FWHM values more than 3 standard deviations
from the median in all IRAC bands; typically this means FWHM$>$2\arcsec.
Most objects located below the [3.6] band limiting magnitude (solid line) in
Figure \ref{fig:agn1} are identified in this way as
extended sources and are likely to be extragalactic objects.
We therefore initially restrict our sample
to objects above the solid line in Figure \ref{fig:agn1}.

\begin{figure}
\epsscale{0.95}
\plotone{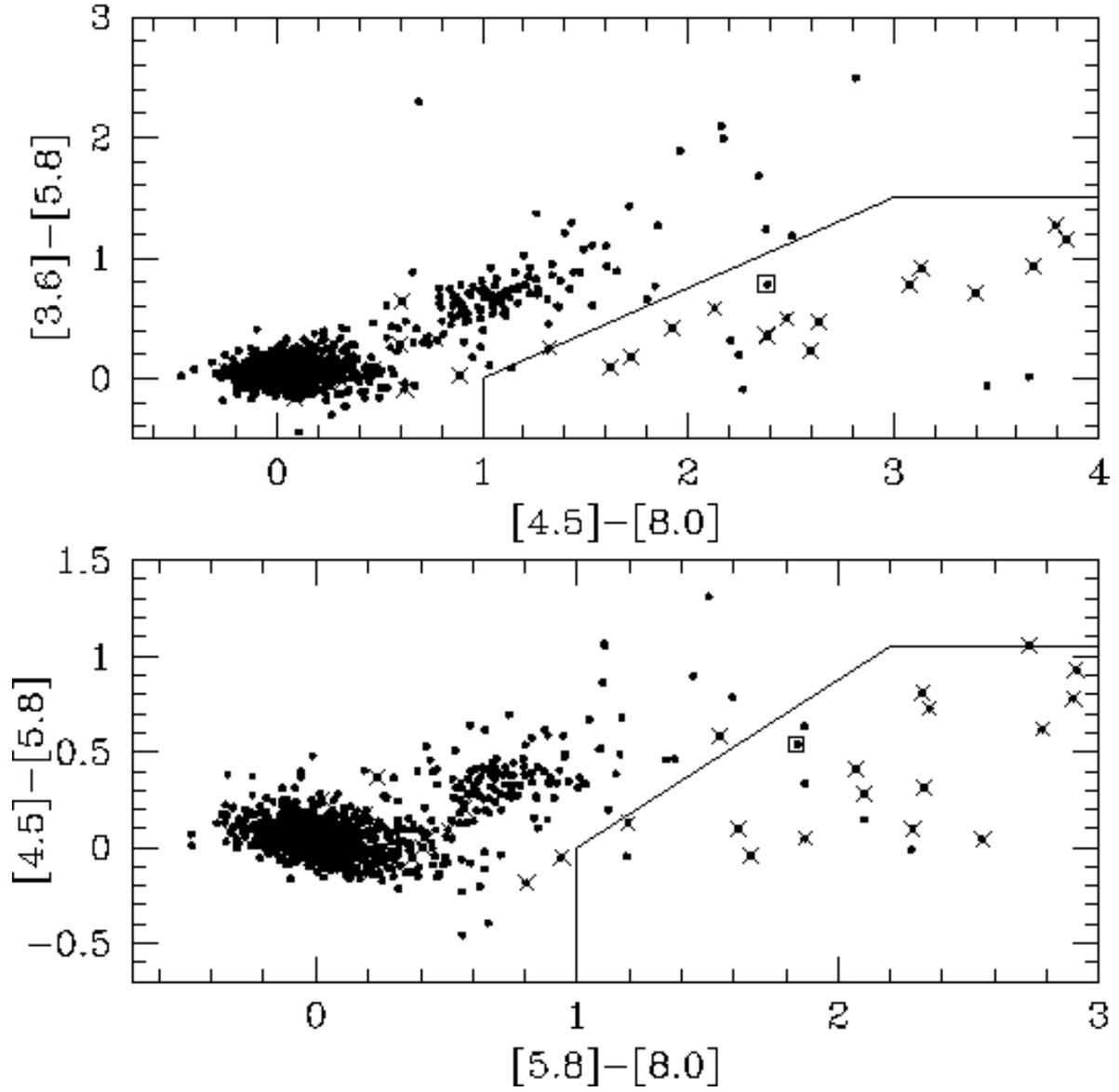}
\caption{IRAC color-color diagrams illustrating the color selection used
to reject PAH-rich galaxies. Symbols are the same as in Figure \ref{fig:agn1}.
Objects located redward from the solid lines are likely to be galaxies
or faint companions with photometry affected by the primary stars of the
stellar system. The open square is the star HD294268 (SO411)}
\label{fig:pah}
\end{figure}

Some galaxies have very red colors because
of strong PAH (polycyclic aromatic hydrocarbon) emission; 
with features at 7.7, 8.6, 11.2, 12.7, 16.4, and 17.1 {\micron} correlated 
with star formation rates \citep{wu05}. 
\cite{gutermuth06} showed that the [4.5 - 5.8] versus [5.8 - 8.0]
and [3.6 - 5.8] versus [4.5 - 8.0] color-color
diagrams can be used to eliminate most of these objects.
As shown in Figure \ref{fig:pah}, objects below and to the right
of the solid lines are likely 
to be PAH-rich galaxies. We find that most of the PAH-rich galaxy candidates 
have extended radial profiles as inferred from point-source fitting.
Visual inspection of the IRAC images also shows 
that many of these candidates are diffuse objects.  In addition to the galaxies,  
other objects with similar colors turn out to be faint companions 
of brighter stars, thus calling into question the accuracy of the photometry.  
Only one object (open square) with a stellar 
profile is located in the PAH-rich galaxies regions; 
this star (HD294268=SO411) has been identified as a young star with spectral 
type F5 \citep{gregorio02}. We therefore eliminate the rest of the objects below 
the solid lines in Figure \ref{fig:pah}, resulting in a sample of 1280 stellar 
sources (hereafter {\it Sample 2}).

About 40\% of the objects in {\it Sample 2}
do not have optical counterparts in the CIDA Equatorial Variability survey; most of these objects are
fainter than the limiting magnitude (V$\sim$19.7) or brighter than the saturation limit
(V$\sim$13.5) of this survey. Some of the objects in {\it Sample 2} also
are located on the  gap in declination of the survey.
We augmented the V optical data set using photometry from \citep{sherry04}
and for the brightest objects from the \citet{kharchenko01} catalog.

\subsection{Additional photometric selection}

\begin{figure}
\epsscale{0.9}
\plotone{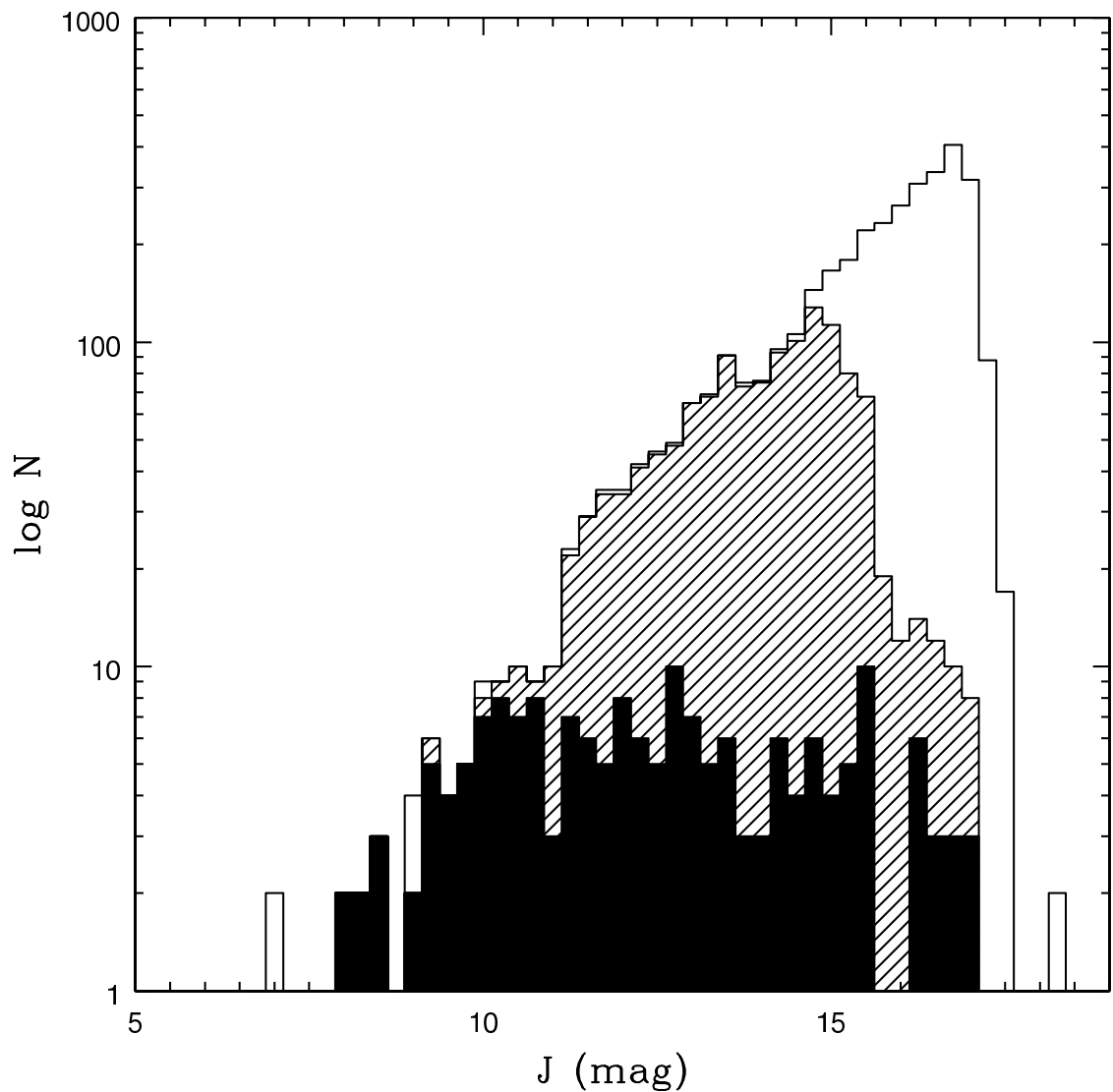}
\caption{2MASS J magnitude distribution: the open histogram represents 
all the 2MASS sources located in the IRAC field. Striped and solid histograms
represent the 2MASS sources detected in IRAC and MIPS, respectively. Our completeness 
limit for sources with [3.6]$<$14.5 (see Figure \ref{fig:agn1}) is  J$\sim$14. 
Most objects with J$<$11 have MIPS detections.}
\label{fig:comp}
\end{figure}

Virtually all the objects in {\it Sample 2} (99\%) have 2MASS counterparts
\citep{cutri03}.
Figure \ref{fig:comp} shows the distributions of J magnitudes for all 2MASS sources in the
IRAC field (open histogram)
and for the 2MASS sources included in {\it Sample 2} (striped histogram).
Comparison of these two histograms indicates that {\it Sample 2} is essentially complete to about
J=14.0.  Assuming a reddening of E(B-V)=0.05 \citep{brown94,bejar99},
a distance of D=440 pc \citep{brown98, hernandez05}
and the 3 Myr isochrones from \citet{baraffe98}, the equivalent completeness limit in mass
is $\sim$ 0.15 $\msun$ (spectral type $\sim$ M5).
For comparison, the distribution of J magnitudes  for our MIPS detections (solid histogram)
is shown in Figure \ref{fig:comp}.  Most of brightest objects (J$<$11; $\sim$1.5 $\msun$)
have 24 {\micron} detections.

{\it Sample 2} includes 214 stars previously found to be members by
other investigators
(hereafter previously known members; see the Appendix \ref{sec:members}). 
Of these objects, the membership of 104 stars were confirmed spectroscopically,
while the other 110 stars were assigned membership based only on optical photometry.
In addition, 105 objects in our sample have been identified as X-ray sources
by \citet{franciosini06}, of which 49 objects have not been identified 
previously as spectroscopic or photometric members. 

\begin{figure}
\epsscale{0.9}
\plotone{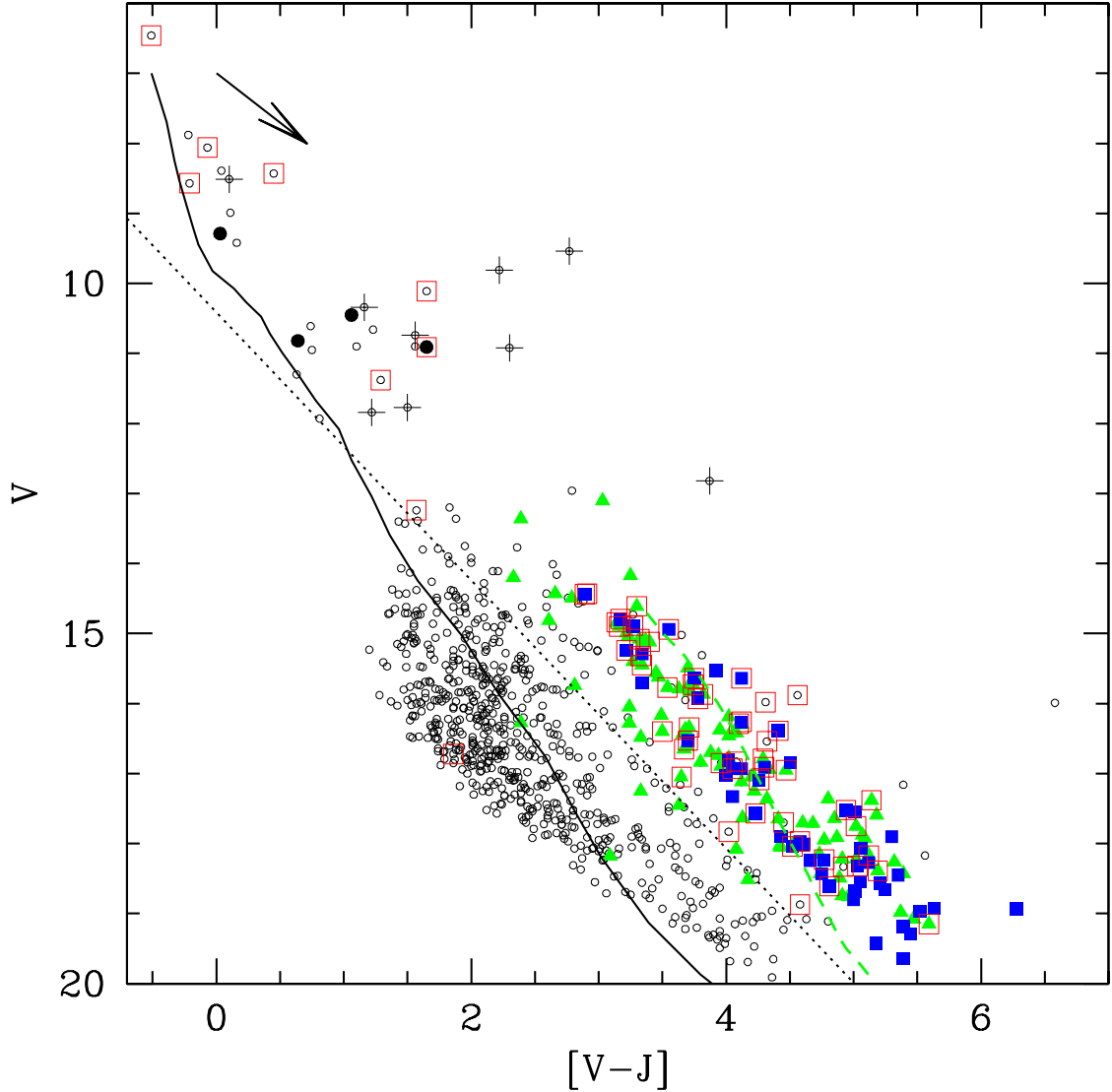}
\caption{V versus V-J color-magnitude diagram, illustrating the 
selection of members of the $\sigma$ Orionis cluster. 
Open circles represent stars in {\it Sample 2}.
Solid symbols represent previous known members: 
photometric members (filled triangles), spectroscopic members (filled squares), 
and members selected by \citet{kharchenko04} combining photometric and astrometric data 
(filled circles). 
Stars with low membership probability combining photometric
and astrometric data are labeled with the plus symbol.
Open squares represent the X-ray sources in the sample
\citep{franciosini06}.
The ZAMS from \citep{sf00} is represented as solid line and 
the 3 Myr isochrone from \citet{baraffe98} is represented as a 
dashed line. The arrow represents the reddening vector (\av=1). 
The membership limit is defined by the dotted line; 
stars located blueward from this limit 
are rejected as likely non-members of the $\sigma$ Orionis cluster 
[See electronic edition of the journal for a color version of this Figure]}
\label{fig:cmd_sel}
\end{figure}

Figure \ref{fig:cmd_sel} shows the V versus V-J color-magnitude
diagram for {\it Sample 2}.
Filled squares (spectroscopic members) and filled triangles (photometric members)
represent previously known members;
both distributions share the same region in Figure \ref{fig:cmd_sel}
suggesting that optical photometric selection can be effective
in identifying low mass members \citep[see also; ][]{kenyon05, burningham05}.
The stars with $3 < V-J < 4$ roughly follow the 3 Myr isochrone of \citet[][dashed curve]{baraffe98}, 
while the redder stars depart significantly from the theoretical
isochrone; this may be due to incompleteness in 
the opacity tables at low temperatures, resulting in theoretical colors that are too blue for 
V-J $>$ 4 \citep{lyra06,baraffe98}.
Filled circles represent the higher mass stars selected by \citet{kharchenko04}
with high membership probabilities ($>$50\%); the plus symbols represent
bright stars with very low probabilities of membership.

The mean of the distribution of previously known cluster members 
can be characterized roughly as straight line in
the V versus V-J diagram, with a standard deviation $\sigma$(V-J)~$= 0.27$ mag.
As a first approximation, we define the photometric region
of probable members as the region above the dotted line in Figure 
\ref{fig:cmd_sel}.  This line represents the lower 3 $\sigma$ deviation 
from the mean distribution line.
The X-ray detected stars \citep{franciosini06}, 
labeled by open squares, are mostly located in the region of
the previously known cluster members (most of these objects were identified previously
as cluster members).  Therefore, X-ray emission can be used
as an additional criterion supporting membership in young stellar groups
\citep[e.g.; ][]{carkner98, preibisch05, franciosini06}.

\begin{figure}
\epsscale{0.9}
\plotone{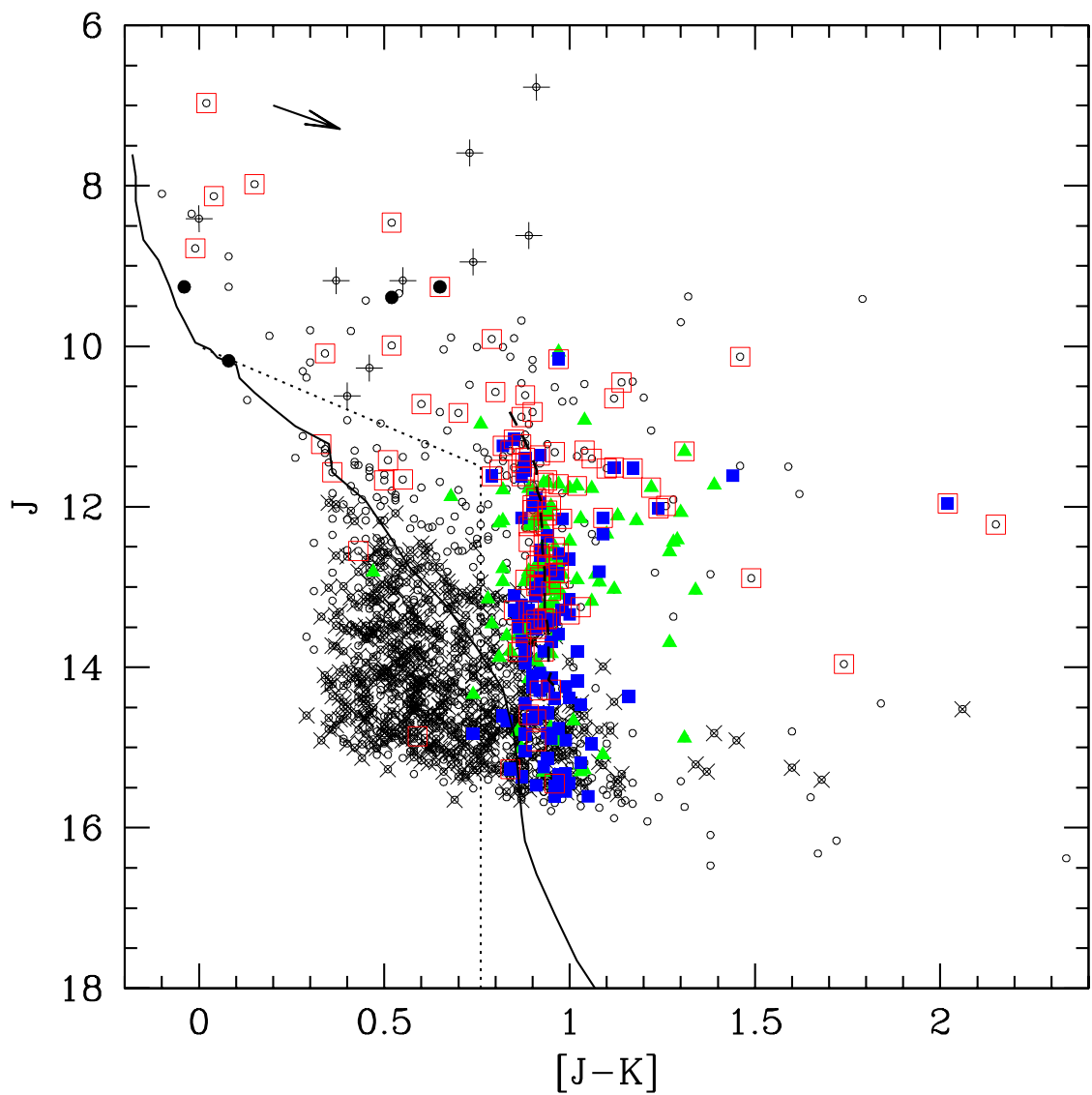}
\caption{J versus J-K color-magnitude diagram, illustrating the 
selection of members of the $\sigma$ Orionis cluster. Symbols are similar 
to Figure \ref{fig:cmd_sel}. Stars rejected as likely non-members in Figure \ref{fig:cmd_sel}
are represented with crosses. The arrow represents the reddening vector (\av=1).
Contamination by non-members are expected
for stars with J$>$13. We reject as non-members of the $\sigma$ Orionis cluster stars
located blueward from the dotted line. Stars selected using only the 
criterion defined in this plot are cataloged as uncertain members
[See electronic edition of the journal for a color version of this Figure]}
\label{fig:cmd_sel2}
\end{figure}

Figure \ref{fig:cmd_sel2} shows the J versus J-K color-magnitude diagram for {\it Sample 2}.
Stars with J~$>$~11.5 have similar J-K colors,
in agreement with the trend expected for the 3 Myr isochrone
from \citet{baraffe98}. For these stars, the distribution of J-K color can be
represented as a Gaussian function centered at J-K=0.94 with $\sigma$(J-K)~=~0.06.
We adopt the border between members and non-members to be the 3 $\sigma$
deviation line (dotted line). For stars with J$<$11.5, we define the region of probable
members by using the X-ray sources in addition to the known members of the cluster.
The crosses represent non-members defined using  Figure \ref{fig:cmd_sel}
(below the dotted line). Some of these non-members, with J$>$13,
appear in the region of probable members in Figure \ref{fig:cmd_sel2},
so some contamination would result from selecting stars using
2MASS colors alone.

Our preliminary member sample consists of stars located redward from dotted 
lines in  Figures \ref{fig:cmd_sel} and \ref{fig:cmd_sel2}, X-ray sources \citep{franciosini06}, 
and previously known members 
(see \S\ref{sec:members}) located redward from the dotted line in  Figure 
\ref{fig:cmd_sel2}. Our preliminary uncertain member sample of the cluster  
includes those objects selected using only the criterion defined in Figure \ref{fig:cmd_sel2},
where the contamination of non-members could be important and additional confirmation is required.

The most important source of contamination are field M dwarfs.  Given the galactic
latitude of the $\sigma$ Orionis cluster (b$\sim$17.3 deg), M giant stars are not expected to
be a significant source of contamination ($<$5\%) in the cluster \citep{bejar99,bejar04}.
We identify 3 possible M giants stars using the 
J-H versus H-K color-color diagram \citep{bessell88},
including them in the uncertain members sample.

\citet{jeffries06} showed that young stars located in the general region
of the cluster consist of two populations with different 
mean ages and distances superimposed along the line of sight: one with an 
age of 3 Myr located at 439 pc (consistent with the $\sigma$ Orionis cluster) and the other 
with 10 Myr located at 326 pc, which could be associated with the Orion 
OB1a association \citep{jeffries06, briceno06}.
However, most of the objects belonging to the older stellar group are located
north-ward from the main population of $\sigma$ Orionis cluster, so that 
the contamination by the older stars in our survey should be small, 
$\lesssim$10\% (Jeffries 2006, private communication). 
In particular, 11 out 75 stars included in our preliminary member sample 
have radial velocities from \citet{jeffries06} consistent for stars in the older
population but inconsistent with those of the
$\sigma$ Orionis cluster. These stars were included, with a note, in the 
uncertain member sample.
Only two stars in our uncertain members sample have radial 
velocity expected of stars in the  $\sigma$ Orionis cluster; we included these 
in our final membership list.

The final catalogs include 336 stars in the member sample and 
133 stars in the uncertain member sample. 
In the member sample, 132 stars have optical photometry from the CIDA survey,
of which 57 stars (43\%) were classified as variable stars using the
method described in \citet{vivas04};
since young stars are expected to exhibit photometric variability 
\citep[e.g.; ][]{cohen76, herbst99, briceno05}, this is 
an additional criterion supporting membership.

Table \ref{tab:members} shows IRAC, MIPS and V photometry data for stars in the 
{\it member sample} according to the criteria discussed above. 
Column 1 shows the internal running identification number in our sample; columns 
2 and 3 are the stellar coordinates; columns 4, 5, 6 and 7 give IRAC magnitudes 
in the bands [3.6], [4.5], [5.8]
and [8.0], respectively; column 8 gives MIPS (24 {\micron}) magnitudes; columns
9 and 10 show the V magnitudes and its sources; column 11 indicates if the star has
been identified as variable in the CIDA survey; column 12 indicates if the star is
a x-ray source \citep{franciosini06}; column 13 shows the disk classification based on
on the IRAC and MIPS analysis (see \S \ref{sec:res}); the last
column gives the references for the {\it previously known members}.
Table \ref{tab:uncert} shows IRAC and MIPS photometry for stars included in the
{\it uncertain member sample}. The information shown in this table in columns from 1 to 8 is
the same as in Table \ref{tab:members}.  The last column shows the disk classification based on
the IRAC and MIPS analysis discussed in the following section.

\section{Results}
\label{sec:res}

\subsection{Infrared color-color diagrams}
\label{res:cc}

\begin{figure}
\epsscale{1.0}
\plotone{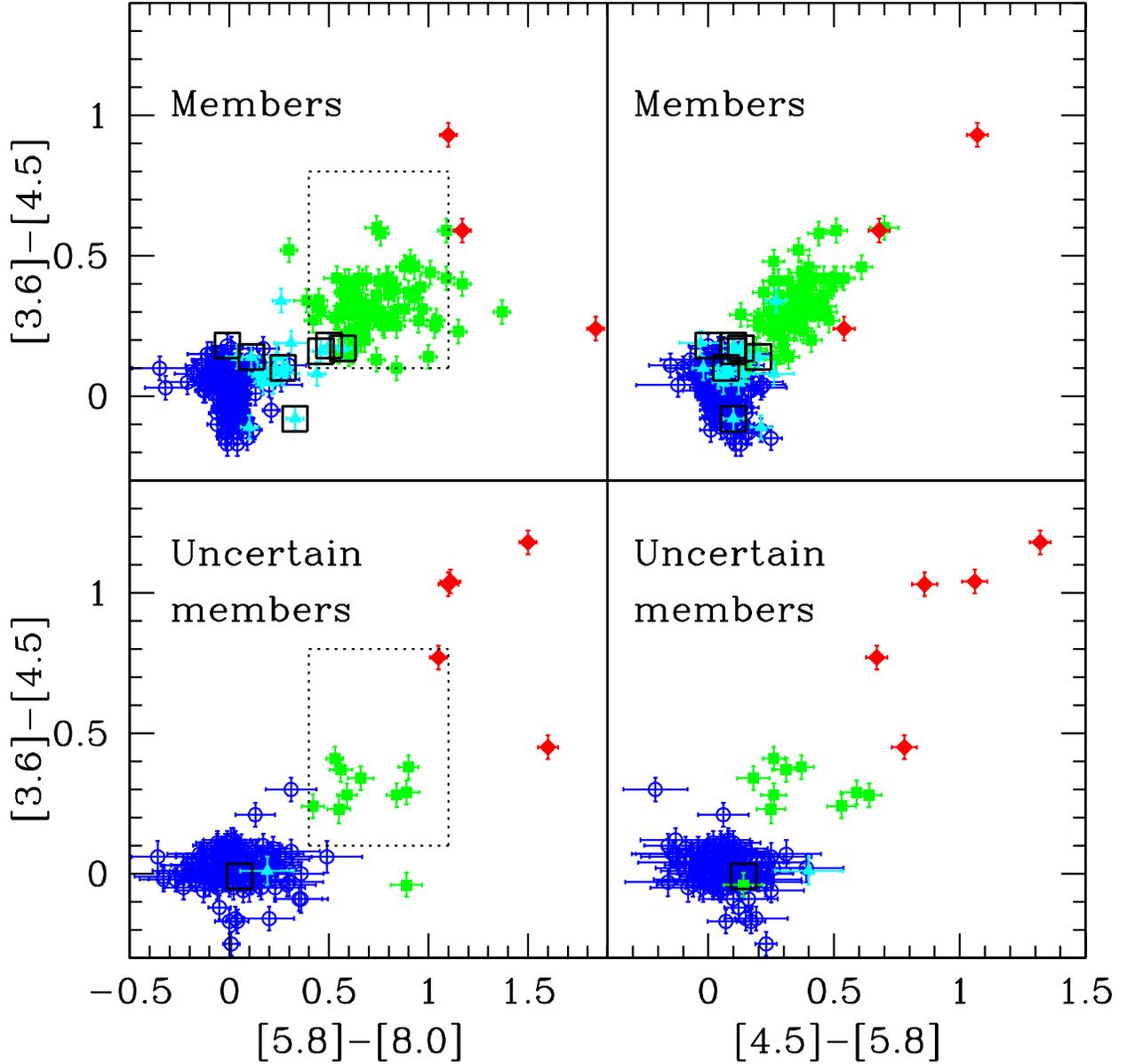}
\caption{IRAC color color diagrams, [5.8]-[8.0] versus [3.6]-[4.5] (left panels) and 
[5.8]-[8.0] versus [4.5]-[5.8] (right panels). Members are plotted in the upper panels 
and uncertain members are plotted in the lower panels. Dashed box represents the loci 
of CTTS with different accretion rates \citep{dalessio05b}. Open circles, filled triangles, 
filled squares and filled diamonds represent  IRAC class III stars, stars with weak IRAC excess, IRAC class II stars 
and IRAC class I candidate 
(classified in \S \ref{res:slope}), respectively.
Objects surrounded by open squares are transition disk candidates (see Figure \ref{fig:mips_cc})
[See electronic edition of the journal for a color version of this Figure]}
\label{fig:irac_cc}
\end{figure}

Figure \ref{fig:irac_cc} shows two IRAC color-color diagrams, [5.8]-[8.0] 
versus [3.6]-[4.5] (left panels) and [4.5]-[5.8] versus [3.6]-[4.5] (right panels),  
for the members (upper panels) and uncertain members (lower panels).
Different symbols show the disk classification based on the IRAC SED slope (\S \ref{res:slope}). 
Particularly, triangles, squares and diamonds represent stars 
with different levels of infrared emission, while circles are non-excess stars.
Symbols surrounded by open squares show objects with small or no excesses in the IRAC
bands but large excesses at 24 {\micron} (see below).
The dashed box in the [3.6] - [4.5] versus [5.8] - [8.0] color-color diagram 
indicates the the region predicted for stars with 
optically-thick disks (\S \ref{res:slope})
by the models of \citet{dalessio05b} 

The upper-left panel of Figure \ref{fig:irac_cc} shows 
that most of stars with significant IRAC excess (squares; hereafter IRAC class II stars) 
have colors in agreement with model colors for optically-thick 
disks. The stars without IRAC excesses (hereafter IRAC class III stars) are located around 
the photospheric region  ([5.8]-[8.0]$\sim$0, [3.5]-[4.5]$\sim$0). 
A few systems are found in the region between the IRAC class III  
and class II stars; these objects with weak IRAC excess (triangles) 
are discussed in more detail in \S \ref{res:slope}.
Two of these objects (the stars SO638 and S0299)  have 
[3.6]-[4.5]$\lesssim$0, while exhibiting significant emission at 
longer wavelengths.  
Three other objects (diamonds) have very large
excesses which may suggest the presence 
of a protostellar envelope (hereafter IRAC class I candidates);
SO457 (IRAS05358-0238), 
SO927 \citep[an emission line object][]{weaver04}, and SO411 (HD294268). 
The last two objects have [3.6]-[4.5] colors expected for class II systems, 
so these objects are more likely to be objects with
optically-thick disks and high 8 {\micron} 
fluxes due to strong silicate emission or rising SEDs at wavelength 
longer than 5.8 {\micron} (\S \ref{res:sed}).

In the upper-right panel of Figure \ref{fig:irac_cc}, 
the [4.5]-[5.8] colors of IRAC class II stars
shows less scatter in comparison with the [5.8]-[8.0] colors. 
There may be some PAH background contamination present in the [8.0] band; 
however, this also could indicate that intrinsic differences in disk flaring
or other properties that may be more easily detected at $8 \mu$m \citep{dalessio06}. 
For example, the IRAC class II stars 
located redward from the dashed box (SO462, SO774 and SO1266) and 
the star S0411 (with [5.8]-[8.0]$>$1.5) in the upper-left panel, 
have a flat or rising SEDs at wavelengths longer 
than 5.8 {\micron} (see \S \ref{res:sed}). 

In the lower panels, nine IRAC class II stars 
are located in the region expected for stars with optically thick disks, 
and thus are likely to be cluster members.
One IRAC class II star (SO1356), located below the model region, 
has an excess at [5.8]-[8.0], but not at [3.6]-4.5] or [4.5]-[5.8]. 
This star was not covered in the MIPS field, so it is not clear if the excess
observed at 8 {\micron} originates from PAH background contamination or from
emission from an outer disk.
The object SO1293 (represented by the triangle) has a 
larger excess at [4.5]-[5.8] than at [5.8]-[8.0];
it is faint (J=15.64) and so could have a longer wavelength excess
below the MIPS detection limit. 

Five uncertain members with very large infrared excesses (IRAC class I candidates), 
are plotted in the IRAC color-color diagrams: SO950, SO336, SO361, S0916 and SO668. 
Because AGN are located in the same region as the class I objects in these diagrams 
\citep{stern05} and are too faint to easily detect spatial extension, additional data 
are needed to confirm these objects as protostellar members.

\begin{figure}
\epsscale{1.0}
\plotone{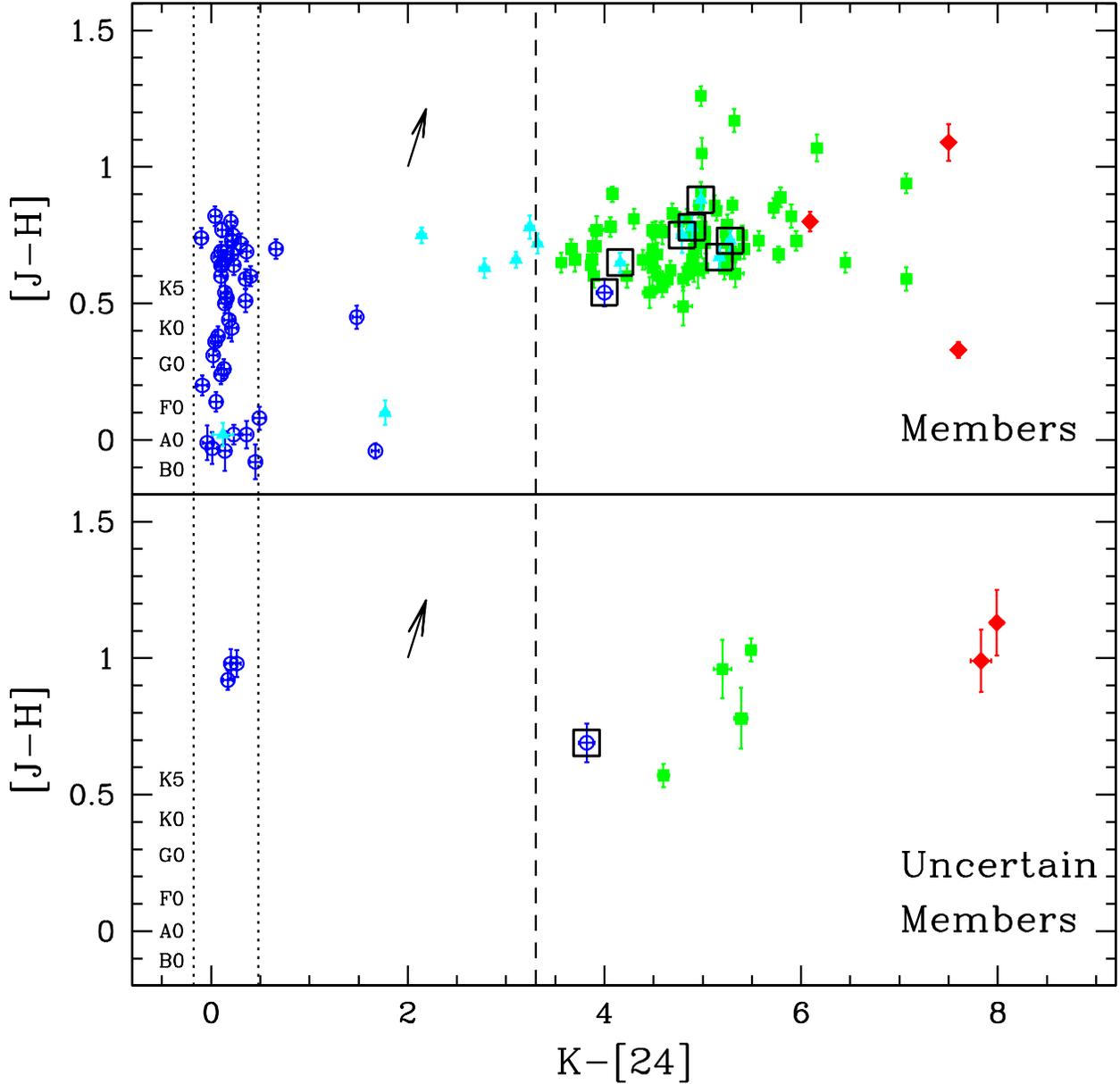}
\caption{[J-H] versus K-[24] color-magnitude diagrams for members (upper panel) and 
uncertain members (lower panel) in the $\sigma$ Orionis cluster. Symbols are similar to Figure 
\ref{fig:irac_cc}. Dotted lines define the locus expected for non emission stars. The dashed line 
at K-[24]=3.3 (the color of the debris disk star $\beta$ Pic) indicates the limit between 
optically thick disk objects and evolved disk objects; transition disk candidates are stars
classified as IRAC class III or evolved disk objects located redward from the dashed line 
(objects surrounded by open squares). In general, the IRAC class I candidates are locate
 redward from the optically thick disk objects.  
The arrow represents the reddening vector (\av=2)
[See electronic edition of the journal for a color version of this Figure]}
\label{fig:mips_cc}
\end{figure}

Figure \ref{fig:mips_cc} shows K-[24] versus J-H color-color diagrams for members (upper panel)
and uncertain members (lower panel). 
We sort the objects into three regions, depending upon the magnitude of their
K-[24] excesses \citep[see also ][]{hernandez06}.  One region consists of 
K-[24]~$<0.5$, corresponding to systems with
undetectable or very uncertain long-wavelength excesses, given photometric errors.
A second region encompasses objects
with K-[24]~$>$3.3, corresponding to optically-thick disk emission
and/or protostellar envelope emission.  Finally, the intermediate-excess region
0.5$<$~K-[24]~$<$ 3.3 is consistent with colors typical of optically-thin
debris disks [the famous $\beta$ Pic debris disk system,
which is one of the debris disk system with largest 24 {\micron} excess known,
has K-[24]$\sim$3.3; \citep[e.g., ][]{gorlova04,rieke05}]. 
 
There are seven members with very large [24] excesses
but low IRAC excesses (symbols surrounded by open squares); 
SO908, SO818, SO1267, SO897, SO299, SO587 and SO1268. 
The last object has no detectable IRAC excess.
These stars are possible ``transition disk'' systems, in which the
inner disk is completely or nearly-completely cleared of small dust
but the outer disk is optically thick \citep[][see \S \ref{res:cc}]{calvet02}.
In the lower panel, the uncertain member SO120 is also a transition disk candidate. 

There are two early type members which are located in 
the intermediate-excess region: SO913 (J-H=-0.04) has no IRAC excess 
and SO956 (J-H=0.10) has a weak IRAC excess (\S \ref{res:slope}). 
These stars are ``debris disk candidates''.
The star SO981 (J-H$\sim$0.45) shows no IRAC excess 
but has a modest 24 {\micron} excess, and thus is a likely debris disk
system around a K early type star.
However, a more detailed study of the debris disk candidates in 
this young stellar cluster is necessary
to clarify the actual nature of their 24 {\micron} excesses, 
since contribution from primordial disk material could be present \citep{hernandez06}.
There are 5 low mass star members (J-H$>$0.5) with weak IRAC excess, in which 
the excesses observed at the IRAC bands and at 24 {\micron} are consistent with 
a overall decrease of infrared emission from the inner and outer disks.

\subsection{IRAC SED slope}
\label{res:slope}

\citet{lada06} showed the utility of characterizing excess infrared
emission using the IRAC SED slope $\alpha$ determined from the [3.6]-[8.0] color, 
where $\alpha$=$dlog[{\lambda}F_\lambda]/dlog[\lambda]$).
In this system, IRAC class I candidates, defined as systems with $\alpha > 0$, are usually
protostars; disk-bearing systems,  having $-2.56 \lesssim \alpha < 0$, 
include IRAC class II stars and objects with weak IRAC excess (the lower limit depends 
upon spectral type and photometric errors, see below); 
and IRAC class III stars have essentially no inner disk emission,
with $\alpha < -2.56$ (again, depending upon spectral type and errors).

\begin{figure}
\epsscale{1.0}
\plotone{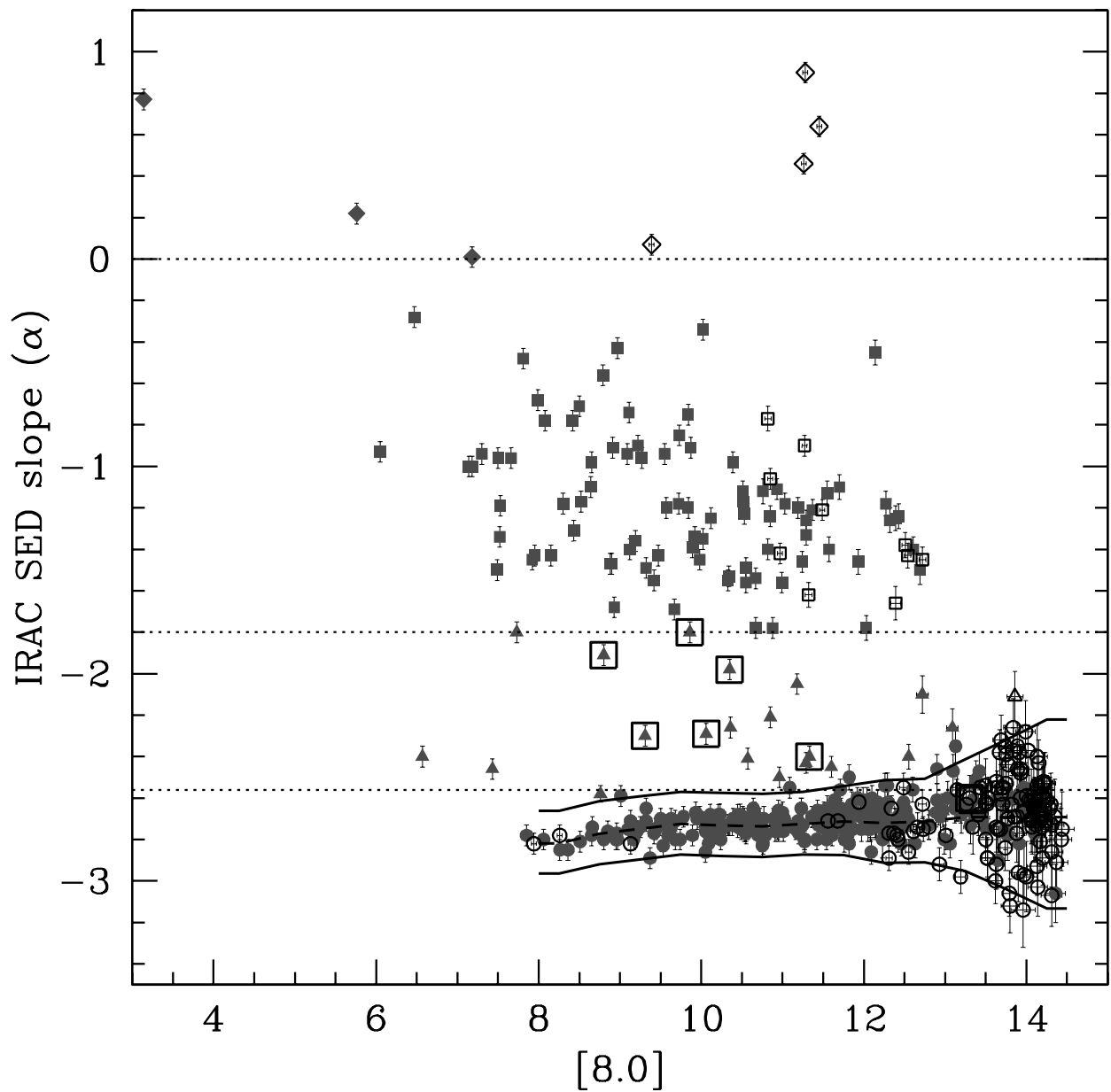}
\caption{IRAC SED slope versus [8.0] 
for members (filled symbols) and uncertain members (open symbols) 
in the $\sigma$ Orionis cluster. Dotted lines represent the limits from 
\citet{lada06} for different objects: IRAC class III stars (circles), 
IRAC class II stars (squares) and evolved disk objects (triangles). 
The dashed line represents the $\alpha$ median for IRAC class III stars
The solid lines represent 3 $\sigma$ values around this median; we adopt the solid upper 
line as the boundary between IRAC class III stars and evolved disk stars. 
The dotted line at $\alpha$=0 indicates 
the boundary between IRAC class II stars and objects with flat or rising SED, class 
I candidates (diamonds). Class III stars and evolved disk objects 
surrounded by open squares are transition disk candidates, since 
they show strong 24 {\micron} excess consistent with having an optically 
thick disk [See electronic edition of the journal for a color version of this Figure]}
\label{fig:alphadef}
\end{figure}

Figure \ref{fig:alphadef} shows $\alpha$
versus the [8.0] magnitude for members (filled symbols) and 
uncertain members (open symbols) of the cluster.
Dashed lines represent the regions of differing levels of infrared
excess as discussed above.  We subdivide the disk-bearing stars into two
categories: IRAC class II stars
(squares, $\alpha$$>$-1.8); and systems with less IRAC infrared excess, 
which \citet{lada06} termed "anemic" disks
and which we call here ``evolved disks" (-2.56 $\lesssim \alpha < -1.8$).

Because the photometric uncertainties in $\alpha$ increase with [8.0],
a single straight line in Figure \ref{fig:alphadef} to separate IRAC class III
stars (in the IRAC wavelength range) and evolved disks is not useful. 
\citet{lada06} avoided this problem by applying their criteria
to separate class III and evolved disks only to the
brighter stars with spectral types M4.5 or earlier
(the number of evolved disks would be overestimated in the lowest
mass bins due to large photometric errors; C. Lada, private communication).
We distinguish between IRAC class III stars (circles) and evolved disk systems (triangles)
using the mean error of $\alpha$ ($\bar{\sigma}$) calculated from the photometric errors 
for stars within 0.5 magnitude bins.  IRAC class III stars are those with excesses, 
if any, smaller than the 3 $\bar{\sigma}$ deviation (solid lines) from the median 
$\alpha$.

\begin{figure}
\epsscale{1.0}
\plotone{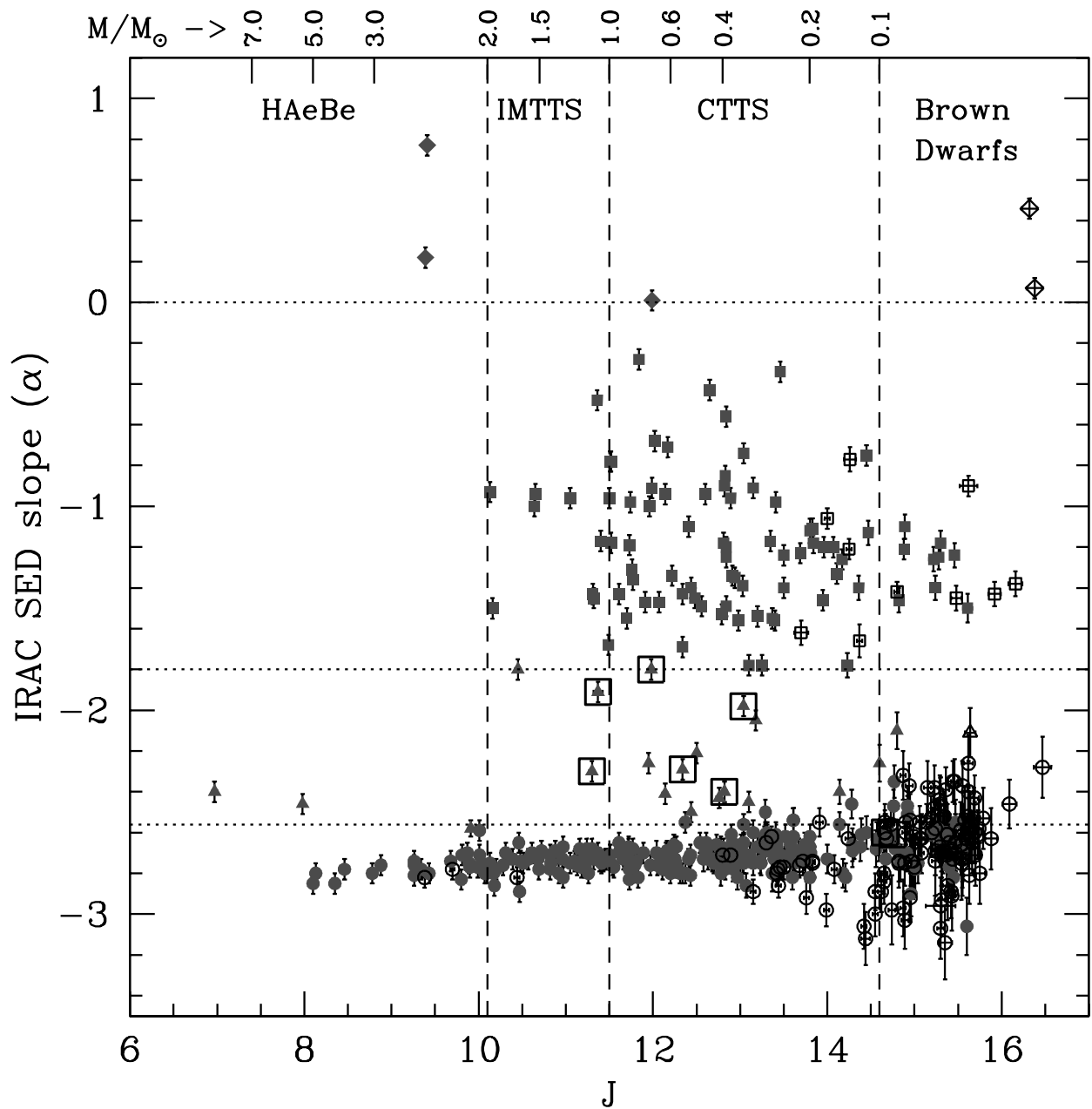}
\caption{IRAC SED slope versus J magnitude 
for stars in the $\sigma$ Orionis cluster. Symbols are as in Figure 
\ref{fig:alphadef}. Vertical dashed lines indicate approximate
mass ranges defined using 3 Myr evolutionary models from \citet[][for $<$1 $\msun$]{baraffe98}
and from \citet[][for $>$1 $\msun$]{sf00}, assuming a distance of 440 pc and a reddening 
of E(B-V)=0.05 [See electronic edition of the journal for a color version of this Figure]}
\label{fig:alphaJ}
\end{figure}

We wish to examine the mass dependence of disk emission, but we do not have
spectral types for most of these stars.  As a proxy for mass, we use the J magnitude,
which is relatively unaffected by disk excess emission, and convert to approximate
mass ranges using the 3 Myr isochrone from \citet{baraffe98} for low mass 
($\leq$1 M$\sun$) stars and \citet{sf00} high mass ($>$1 M$\sun$) stars,
assuming a distance of 440 pc and E(B-V)=0.05.
Figure \ref{fig:alphaJ} shows $\alpha$ versus J, with
the regions expected for objects in different ranges of mass: 
HAeBe range ($>$2 M$\sun$), Intermediate Mass TTS (IMTTS) range (1.0-2.0 M$\sun$),
TTS range (0.1-1.0 M$\sun$) and Brown dwarfs range ($<$0.1 M$\sun$).
 Most of the objects with
optically thick and evolved disks are located 
in the mass range of T Tauri stars. 

Using the results from Figure \ref{fig:alphaJ}, we derive
the fractions of stars with optically-thick disks (filled circles + dotted line) 
and all disk-bearing stars (open circles + dashed line) versus J magnitude,
with corresponding mass ranges indicated (see also Table \ref{tab:frac}). 
The lowest fraction of disks is observed in the highest mass stars (HAeBe range), 
in agreement with  the fraction calculated for the overall population of  OB1b sub-association 
\citep{hernandez05}. There is a marginal evidence that the disk fraction declines 
toward lower masses (into the brown dwarf range), in agreement with results from \citet{lada06}
for the young (2-3 Myr) stellar group IC 348. Nevertheless, including the error bars, 
the fraction is also consistent with no mass dependence toward lower masses. 
The fraction of thick disks in the TTS mass range agrees with the result from 
\citet[][ 33$\pm$6 \%; ]{oliveira06} for low mass stars.

Only 15$\pm$5\% of disk-bearing stars in the TTS mass range are disk evolved objects, 
approximately in agreement with results from the 4 Myr-old cluster
Tr 37 \citep[10\%; ][]{aurora06}, although with better statistics and higher
photometric accuracy.  If we assume a steady-state evolution of disks, and also
assume that the evolved disks are going through a phase of clearing of small dust
particles in the inner disk, then our observed frequency suggests a timescale
of inner disk clearing $\sim 0.15 \times 3$~Myr~$\sim 0.45$~Myr.

\subsection {Disk diversity: SEDs}
\label{res:sed}

Figure \ref{fig:sed} shows SEDs (data from IRAC, MIPS, 2MASS and optical B, V, R 
and I magnitudes when are available) for selected members.
The first 5 rows show the SEDs of stars in the mass range of TTS (0.1-1.0 M$\sun$), 
while the last row shows the SEDs of stars in the mass range of HAeBe stars ($>$2 M$\sun$).
SEDs (solid lines) are normalized to the J band and are sorted by different levels 
of IR excesses indicating differences in the disks. Each panel is labeled with the 
internal running identification number and the classification from the IRAC SED slope analysis
(\S \ref{res:slope}).  
The dotted line represents the median SED (Table \ref{tab:sed_ctts}) of 
optically thick disk stars in the mass 
range of TTS defined in Figure \ref{fig:alphaJ} (hereafter, optically thick disk median SED); 
error bars denote the quartiles of the distribution. This typical disk emission in the $\sigma$ 
Orionis cluster is lower than the median in Taurus \citep[][dot-dashed line 
in the first upper-left panel]{dalessio99}, which has an age of $\sim$1-2 Myr, indicating 
differences due to evolutionary effects such as grain growth and/or settling to the disk
midplane \citep[see also ][]{aurora06,lada06}, or lower accretion-rate \citep{dalessio06}. 
The dashed lines represent the median photospheric SED (Table \ref{tab:sed_wtts}) 
in the T Tauri mass range.

The first row of Figure \ref{fig:sed} 
shows the SEDs of evolved disk objects (SO905, SO1057, SO759 
and SO587) which exhibit modest infrared disk emission. 
The lower level of infrared excess emission could be 
explained by decreasing the height of the irradiation surface in the disks due to 
a higher degree of settling, which diminishes the degree of flaring, thus producing flatter 
disk structures \citep{dalessio06}.   
The second row of SEDs (SO120, SO1268, SO299 and SO897) 
shows some transition disk candidates,
which exhibit little or no excess in 
IRAC bands but 24 {\micron} emission similar to optically thick disks.
This indicates a removal of small dust particles in the inner disks but not depletion 
or removal in the outer disk.
In the next rows, the stars SO818, SO1267, SO908, SO1156,
SO540 and SO1266 show somewhat weaker IRAC excesses.
The stars SO73, SO927, SO1154, SO1153, SO457, and SO668 show larger 
infrared excesses than the optically thick median, which might 
indicate more highly flared disk structure \citep{dalessio06, dullemond05}, 
edge-on disks,  or a contribution from an envelope.
SO927 has strong emission in the H$\alpha$ \citep{weaver04}. 
SO1153 (V510 Ori; HH 444) and SO1154 (HH 445) are Herbig-Haro objects 
producing collimated outflows \citep{andrews04}.  Finally, 
SO457 has been identified by \citet{oliveira04} as a class I object. 

In the last row of Figure \ref{fig:sed}, SEDs of earlier-type members
are compared with photospheres of stars with
similar spectral types \citep[long dashed line ][]{kh95}. 
The star SO724 \citep[B2; ][]{guetter81} shows small excess at 8 {\micron} but 
has no excess at 24 \micron, this star was classified in \S\ref{res:slope} as 
an evolved disk object possibly due to a background PAH contamination at 8 \micron.
The stars SO913 \citep[B9.5; ][]{guetter81} and SO956 
\citep[A8; ][]{guetter81} show small or no excesses in the IRAC bands but have
significant excesses at 24 {\micron}, consistent with being debris disk stars.
Finally, the star SO411, classified as class I candidate in \S \ref{res:slope},
shows small excesses at wavelengths shorter than 5.8 {\micron} and a rising SED 
at longer wavelengths. This emission line star, with spectral type 
F5 \citep{gregorio02}, could be an accreting transition disk object 
in the HAeBe mass range. This object alternatively could be considered to be an IMTTS,
as F5 is the usual transition spectral type between HAeBe and IMTTS \citep[see ][]{hernandez04}. 

Most of the disk classifications using the IRAC SED slope (\S \ref{res:slope})
agree with the results from analyzing SEDs.  However, a few objects were reclassified
taking into account the overall SEDs; these objects are labeled in Tables 
\ref{tab:members} and \ref{tab:uncert}.

\subsection {Disk evolution}

Figure \ref{fig:disk_evol} shows the fraction of stars in the TTS mass range 
with near-infrared disk emission in different stellar groups, as a function 
of age \citep{hernandez05, haisch01}. Recent {\em Spitzer} results 
for the young stellar clusters NGC7129 \citep{gutermuth04}, Taurus \citep{hartmann05c}, 
$\eta$ Chameleontis \citep{megeath05a}, Tr37 and NGC7160 \citep{aurora06}, 
IC348 \citep{lada06}, and Upper Scorpius \citep{carpenter06}
have been included in this plot.  
The disk fractions shown in Figure \ref{fig:disk_evol}
include stars with optically thick and evolved disks 
(disk-bearing stars) detected mostly using IRAC photometry. 
The $\sigma$ Orionis cluster (filled square) 
follows (within the scatter of the distribution) the disk 
fraction trend outlined by in other stellar groups 
\citep{haisch01,hillenbrand06}.

Aside from the decrease 
in the disk fraction with age, the amount of 
infrared disk emission also decreases with age. Figure \ref{fig:slopehist} 
shows the distribution of $\alpha$ for low mass stars in the stellar
groups Taurus \citep{hartmann05c}, IC348 \citep{lada06}, 
$\sigma$ Orionis (this work) and Tr 37 \citep{aurora06}. 
As in Figure \ref{fig:sed}, the youngest group, Taurus, has a larger population of stars 
with robust optically thick disks ($\alpha$ $>$ -1). 
In contrast, in the more evolved stellar groups most disk-bearing 
stars  have an $\alpha$ $<$-1, indicating a 
reduction in the height of the disk photosphere (less disk
flaring) in the inner regions.

The Taurus sample suggests a relatively clear break between stars without
IRAC excesses and optically-thick disk systems 
\citep{hartmann05c}, although the sample size is relatively small.
The $\sigma$ Orionis sample also suggests a small fraction of objects with
intermediate or evolved disk excesses, as discussed previously.  
The evolved disk fraction is less certain in Tr 37, due to the larger
photometric errors for that more distant cluster.
The IC 348 sample suggests a larger fraction of evolved disks than
seen in $\sigma$ Ori; the reality of this possible difference requires
further study.

In contrast to IC 348, the $\sigma$ Orionis and  Tr 37 clusters include 
high mass stars located in their center (O9 in $\sigma$ Orionis cluster 
and O7 in Tr37); however, the medians   
of $\alpha$ are similar in these stellar groups (Figure \ref{fig:slopehist}). 
This suggests that external photoevaporation in  $\sigma$ Orionis and  Tr 37
may be not very important for disk dissipation.

\subsection{Spatial distribution of the $\sigma$ Orionis cluster}

Figure \ref{fig:frac1} shows the spatial distribution of members
with and without infrared excesses.  As shown in the upper right panel,
the space density has a peak with a FWHM of 0.25$\degr$ (1.9 pc).
The upper-left  panel shows the ratio of systems with
disks to the total members in radial distance bins 
defined by the semi-circles East-West versus the radial distance from 
the center of the cluster. Solid circles (+dotted line) represent the fraction of thick disks, 
while open circles (+dashed line) represent the total fraction of disk-bearing stars,
defined in  \S \ref{res:sed}.
The error bars represents the statistical ${N}^{-1/2}$ error in our disk fractions, 
where N is the number of stars with disks. 
Similarly, the lower-right panel shows the fraction 
of disks in radial distance bins defined by the semi-circles North-South. 
In contrast to the result of \citet{oliveira06}, using K and L band observations, there is 
evidence for a higher disk fraction near the cluster center.
However, we cannot rule out a small, relatively uniform contamination of non-members 
which could also produce this effect.

There is a slight increase in the disk fraction  in the north-east direction; 
in the direction toward NGC2024 \citep[age $<$ 1 Myr, D$\sim$440 pc; ][]{levine06,eisner03}, 
located at $\sim$1.0 deg from the $\sigma$ Orionis center; this fraction could be affected by 
disk-bearing members of this younger stellar group. Since NGC2024 is centrally concentred,
the fraction could be also affected by stars extends toward the Horsehead nebula 
(located at $\sim$0.6 deg from the $\sigma$ Orionis center), which also
shows several class II objects in separate {\em Spitzer} maps \citep{megeath05b}.
Spectroscopic studies are necessary to test whether the disk fractions in the outer regions 
are affected by non-members.

\subsection{Disk and X-ray emission}
\label{res:xray}

A sub-sample of 81 stars have X-ray luminosities from \citet{franciosini06}, including
8 evolved disks, 21 optically thick disks and 52 class III stars. 
Figure \ref{fig:alphaX} shows the cumulative distribution of class III stars
and stars with optically thick disks as a function of X-ray luminosity. The significance 
level of 64\% in a Kolmogorov-Smirnov (KS) test indicates that class III stars 
have similar X-ray luminosities as stars with disks.
This suggests that the presence of optically thick disks have small or no influence
on X-ray emission, such as might be caused by an accretion shock,
in agreement with the results from \citet{franciosini06}.

\subsection{Contamination expected in the uncertain members sample}

Stars selected as uncertain members are relatively faint (J$>$13.0).
From 133 stars in this sample, only  11 objects have IRAC SED slopes 
consistent with the presence of disks (10 thick disk + 1 evolved disks).  
So, the fraction of disk-bearing stars for the uncertain members (8.3$\pm$2.5 \%) is very low 
compared with the fraction of disk-bearing stars expected for 
members with J$>$13 in the $\sigma$ Orionis cluster (31.3$\pm$4.8). Assuming that 
the source of non-members are class III stars and the disk-bearing 
stars are actual members of the cluster, we can expect that at least 84 
class III stars ($\sim$70 \% of the uncertain member sample) do not belong 
to the $\sigma$ Orionis cluster and are likely to be field M dwarfs or 
other sources of contaminations. Additional 
spectroscopic studies are essential to determine membership for this sample.

\section{Conclusions}
\label{sec:conc}
 
We have used the IRAC and MIPS instruments on board the {\em Spitzer}
Space Telescope to study the frequencies and properties of disks around 
336 members of the 3 Myr old $\sigma$ Orionis cluster, with results for another
133 objects of uncertain membership. 
We confirm and strongly improve upon previous results showing that disk
fractions are much lower around A-type and intermediate-mass T Tauri stars,
implying that disk evolution is much more rapid for stars with masses
$> 1 \msun$.  
In particular, we find that only a few percent of all A-F
stars have optically thick disks, with perhaps another $\sim 15$~\% having
detectable optically-thin, probable debris disks. This fraction of 
probable debris disk is lower that those found in the older Orion OB1b (38\%)
and OB1a (46\%) sub-associations, suggesting that $\sigma$
Orionis cluster  could be in a phase between the clearing of the 
primordial disk and the formation of large icy objects, which produce the
second-generation dust observed in debris disk systems \citep{hernandez06}.
The disk frequency $\sim 30 - 35$\% we find for T Tauri stars is reasonably
consistent with studies of other populations, perhaps a bit lower.
We also find no evidence that the disk fraction decreases
significantly among the highest-mass brown dwarfs.  
Overall, disk frequencies 
follows the overall trend of decreasing disk frequency with increasing age 
observed in other young stellar populations. 

About 15\% of the T Tauri 
disk systems exhibit weak IRAC excesses inconsistent
with simple optically-thick disk models.  If these systems are in the
process of clearing small dust particles from their inner disks, the observed
fraction suggests that the timescale for this evolution is $\lesssim 0.5$~Myr.
We also find that the infrared excesses among optically-thick disks 
in $\sigma$ Ori are systematically lower than the excesses observed
from younger Taurus molecular cloud stars.  This result confirms and strongly extends
previous results indicating an overall decrease in the level of emission of optically-thick
disks with age, which may indicate an overall settling of dust particles to the
disk midplanes, or grain growth, or both.
The populous nature and relatively near distance of the $\sigma$ Orionis cluster
allows us to improve the statistics of these correlations, providing
an important step toward a full characterization of protoplanetary
disk evolution in clustered environments, in which most stars form.
Finally, we do not find correlation between X ray luminosity and the 
disk infrared excess, suggesting that the presence of optically thick disks 
have small or not influence  on the X ray emission.

\acknowledgements
Thanks to R.D. Jeffries, D. Stern and C. Lada for insightful 
communications. This publication makes use of data products from 
the CIDA Equatorial Variability Survey, obtained with
he J. Stock telescope at the Venezuela National Astronomical 
Observatory, which is operated by CIDA for the Ministerio de 
Ciencia y Tecnolog{\'\i}a of Venezuela, 
and  Two Micron All Sky Survey, which is a joint project of the University 
of Massachusetts and the Infrared Processing and Analysis Center/California 
Institute of Technology. 
This work is based on observations
made with the {\em Spitzer Space Telescope} (GO-1 0037 and GO-1 0058), which is 
operated by the Jet Propulsion Laboratory, California Institute of Technology under
a contract with NASA. Support for this work was provided by University of Michigan 
and NASA Grant NAG5-13210.

\appendix
\section{Previous membership studies}
\label{sec:members}

The $\sigma$ Orionis cluster has been the subject of several photometric 
and spectroscopic studies to identify members of its population..  
Analyzing the photometric and kinematic properties of the bright population
in the cluster, \citet{kharchenko04} determined an angular size of  0.40 
degrees and identified 7 stars with photometric and kinematic membership probabilities larger than 50\%.
Since the cluster is reasonably close, relatively young and has low reddening,
many of the previous studies have focused on identifying low mass stars and substellar objects.
Adopting an age interval of 2-7 Myr and the Hipparcos distance of the star $\sigma$ Ori 
(352$^{+166}_{-85}$ pc), \citet{bejar99,bejar01,bejar04} presented several lists 
of low mass stars, brown dwarfs and free-floating planets belonging to the $\sigma$ Orionis cluster,
selected from optical and NIR color-magnitude diagrams.
Adopting an age of 2.5 Myr and the distance calculated for OB1b subassociation (440 pc), 
\citet{sherry04} reported photometric membership 
for a sample of  low mass stars (193 objects) based on the V versus V-I color-magnitude 
diagram. \citet{sherry04} found that the cluster has $\sim$160 members in the mass range 
from 0.2 $\msun$ to 1.0 $\msun$. Adopting an age of 3 Myr and a distance of 
350 pc, \citet{scholz04} reported photometric membership 
for 135 candidates of very low mass stars ($<$0.4 $\msun$) using the color magnitude 
diagrams, I versus I-J, and photometric variability.

\citet{zapatero02} presented intermediate and low resolution spectra
including the H$\alpha$ and \ion{Li}{1} $\lambda$6708{\AA} lines and  confirmed  
25 low mass stars and 2 brown dwarfs as members of the cluster.
\citet{barrado03} confirmed 51 low mass stars and brown dwarfs as members
using low resolution spectroscopic analysis of forbidden lines, as well as
\ion{Li}{1} $\lambda$6708{\AA} and H$\alpha$ lines.
\citet{kenyon05} presented an intermediate-resolution spectroscopic 
study of 76 photometrically selected stars ranging in mass from 0.04 $\msun$ to 0.3 $\msun$; 
57 stars were confirmed as members on the basis that they exhibit \ion{Li}{1} $\lambda$6708{\AA} in absorption,
a weak \ion{Na}{1} doublet (8183,8195 \AA) indicating low surface gravity consistent with PMS status, and 
radial velocities expected for the kinematic properties of the cluster. 
\citet{burningham05} confirmed as members 38 low-mass stars 
using high-resolution spectroscopic analysis  of \ion{Na}{1} doublet.
\citet{kenyon05} and \citet{burningham05} showed that optical photometric 
selection alone is very effective to identify low mass members belonging to the cluster,
with small contamination by non-members for stars with I$<$17 magnitudes.

Recently, \citet{oliveira06} compiled 59 spectroscopic confirmed members 
of the cluster and studied the circumstellar disks of the sample using NIR data. 
\citet{franciosini06} detect 175 X-ray sources down to the substellar limit. Half of 
this sample could be new members of the cluster based on the X-ray emission as indicator 
of youth.

\clearpage

\begin{deluxetable}{llllllllllllll}
\rotate
\tabletypesize{\tiny}
\tablewidth{0pt}
\tablecaption{Members of the $\sigma$ Orionis Cluster \label{tab:members}}
\tablehead{
\colhead{SO} & \colhead{RA(2000)} & \colhead{DEC(2000)} & \colhead{[3.6]} & \colhead{[4.5]} & \colhead{[5.8]} & \colhead{[8.0]} & \colhead{[24.0]} & \colhead{V} & \colhead{Ref} & \colhead{Var} & \colhead{X-ray} & \colhead{Disk} & \colhead{Known} \\
\colhead{ID} & \colhead{deg} & \colhead{deg} & \colhead{mag} & \colhead{mag} & \colhead{mag} & \colhead{mag} & \colhead{mag} & \colhead{mag} & \colhead{} & \colhead{ } & \colhead{source} & \colhead{type} & \colhead{member} \\
}
\startdata
  9 & 84.32781 & -2.67275 & 10.99 $\pm$  0.03 & 10.97 $\pm$  0.03 & 10.80 $\pm$  0.03 & 10.92 $\pm$  0.03 & \nodata           & 15.40 $\pm$  0.05 &    q &    0 & \nodata &  	III & 9\\
  27 & 84.34599 & -2.54623 & 13.00 $\pm$  0.03 & 12.94 $\pm$  0.03 & 12.86 $\pm$  0.04 & 12.99 $\pm$  0.05 & \nodata           & 18.42 $\pm$  0.02 &    9 &\nodata & \nodata & III & \nodata\\
  41 & 84.35401 & -2.86730 & 10.30 $\pm$  0.03 & 10.31 $\pm$  0.03 & 10.28 $\pm$  0.03 & 10.25 $\pm$  0.03 & \nodata           & \nodata &\nodata &\nodata & \nodata &  	III & \nodata\\
  52 & 84.36209 & -2.36497 & 10.54 $\pm$  0.03 & 10.56 $\pm$  0.03 & 10.50 $\pm$  0.03 & 10.49 $\pm$  0.03 & \nodata           & 13.77 $\pm$  0.08 &    q &    0 & \nodata &  	III & \nodata\\
  59 & 84.36683 & -2.60175 & 12.65 $\pm$  0.03 & 12.62 $\pm$  0.03 & 12.51 $\pm$  0.04 & 12.52 $\pm$  0.04 & \nodata           & 17.64 $\pm$  0.05 &    q &    0 & \nodata &  	III & \nodata\\
  60 & 84.36792 & -2.40508 & 12.87 $\pm$  0.03 & 12.82 $\pm$  0.03 & 12.82 $\pm$  0.04 & 12.80 $\pm$  0.05 & \nodata           & 18.08 $\pm$  0.11 &    q &    0 & \nodata &  	III & 9\\
  73 & 84.37888 & -2.39521 & 10.90 $\pm$  0.03 & 10.44 $\pm$  0.03 & 10.04 $\pm$  0.03 &  9.11 $\pm$  0.03 &  5.64 $\pm$  0.03 & 16.28 $\pm$  0.10 &    q &    1 & \nodata &  	II & 9\\
  74 & 84.37941 & -2.52922 &  9.46 $\pm$  0.03 &  9.39 $\pm$  0.03 &  9.30 $\pm$  0.03 &  9.32 $\pm$  0.03 &  9.19 $\pm$  0.04 & \nodata &\nodata &\nodata & \nodata &  	III & \nodata\\
  77 & 84.38132 & -2.40748 & 10.99 $\pm$  0.03 & 10.98 $\pm$  0.03 & 10.93 $\pm$  0.03 & 10.93 $\pm$  0.03 & \nodata           & 15.83 $\pm$  0.11 &    q &    1 & \nodata &  	III & 9\\
  82 & 84.38269 & -2.75514 &  9.24 $\pm$  0.03 &  9.21 $\pm$  0.03 &  9.16 $\pm$  0.03 &  9.20 $\pm$  0.03 &  8.95 $\pm$  0.04 &  9.42 $\pm$  0.02 &\nodata &\nodata & \nodata & III &\nodata \\
  92 & 84.38811 & -2.89332 & 10.59 $\pm$  0.03 & 10.74 $\pm$  0.03 & 10.64 $\pm$  0.03 & 10.55 $\pm$  0.03 & \nodata           & 14.99 $\pm$  0.04 &    q &    0 & \nodata &  	III & 9\\
 115 & 84.40181 & -2.24377 & 10.74 $\pm$  0.03 & 10.72 $\pm$  0.03 & 10.67 $\pm$  0.03 & 10.70 $\pm$  0.03 & \nodata           & 15.49 $\pm$  0.10 &    q &    0 & \nodata &  	III & 9\\
 116 & 84.40192 & -2.69906 & 14.19 $\pm$  0.03 & 14.02 $\pm$  0.03 & 13.98 $\pm$  0.06 & 13.97 $\pm$  0.09 & \nodata           & \nodata &\nodata &\nodata & \nodata &  	III & 1,2,3,6\\
 117 & 84.40270 & -2.56669 & 11.82 $\pm$  0.03 & 11.83 $\pm$  0.03 & 11.81 $\pm$  0.03 & 11.76 $\pm$  0.04 & \nodata           & 17.12 $\pm$  0.05 &    q &    0 & \nodata &  	III & 9\\
 123 & 84.40760 & -2.76226 & 11.51 $\pm$  0.03 & 11.47 $\pm$  0.03 & 11.40 $\pm$  0.03 & 11.44 $\pm$  0.03 & \nodata           & 16.37 $\pm$  0.05 &    q &    1 & \nodata &  	III & 9\\
 124 & 84.40766 & -2.83598 &  9.49 $\pm$  0.03 &  9.54 $\pm$  0.03 &  9.45 $\pm$  0.03 &  9.44 $\pm$  0.03 &  9.65 $\pm$  0.05 & \nodata &\nodata &\nodata & \nodata &  	III & \nodata\\
 147 & 84.42406 & -2.48557 &  9.51 $\pm$  0.03 &  9.50 $\pm$  0.03 &  9.47 $\pm$  0.03 &  9.53 $\pm$  0.03 &  9.59 $\pm$  0.04 & 10.90 $\pm$  0.07 &    7 &\nodata & \nodata &  	III &\nodata \\
 164 & 84.43706 & -2.49920 &  9.91 $\pm$  0.03 &  9.90 $\pm$  0.03 &  9.91 $\pm$  0.03 &  9.89 $\pm$  0.03 &  9.80 $\pm$  0.05 & 10.95 $\pm$  0.09 &    7 &\nodata & \nodata &  	III &\nodata \\
 165 & 84.43856 & -2.48111 & 12.65 $\pm$  0.03 & 12.58 $\pm$  0.03 & 12.55 $\pm$  0.04 & 12.49 $\pm$  0.04 & \nodata           & 18.76 $\pm$  0.13 &    9 &\nodata & \nodata &  	III & 9\\
 168 & 84.43889 & -2.73678 &  8.96 $\pm$  0.03 &  9.02 $\pm$  0.03 &  8.95 $\pm$  0.03 &  8.95 $\pm$  0.03 &  8.94 $\pm$  0.04 & 10.66 $\pm$  0.07 &    7 &\nodata & \nodata &  	III &\nodata \\
 194 & 84.45671 & -2.60506 & 10.12 $\pm$  0.03 & 10.17 $\pm$  0.03 & 10.06 $\pm$  0.03 & 10.10 $\pm$  0.03 & 10.03 $\pm$  0.06 & \nodata &\nodata &\nodata & \nodata &  	III & \nodata\\
 209 & 84.46288 & -2.43542 & 13.64 $\pm$  0.03 & 13.53 $\pm$  0.03 & 13.68 $\pm$  0.05 & 13.38 $\pm$  0.07 & \nodata           & \nodata &\nodata &\nodata & \nodata &  	III & 2,3\\
 214 & 84.46497 & -2.59045 & 10.81 $\pm$  0.03 & 10.80 $\pm$  0.03 & 10.79 $\pm$  0.03 & 10.72 $\pm$  0.03 & \nodata           & 15.64 $\pm$  0.01 &    9 &\nodata &  1 &  	III & 8,9\\
 219 & 84.46685 & -2.60126 & 13.92 $\pm$  0.03 & 13.81 $\pm$  0.03 & 13.73 $\pm$  0.05 & 13.70 $\pm$  0.07 & \nodata           & \nodata &\nodata &\nodata & 0 &  	III & 6\\
 220 & 84.46747 & -2.56052 & 11.85 $\pm$  0.03 & 11.85 $\pm$  0.03 & 11.78 $\pm$  0.03 & 11.77 $\pm$  0.04 & \nodata           & 16.89 $\pm$  0.03 &    9 &\nodata & 0 &  	III & 9,12\\
 229 & 84.47091 & -2.55951 &  9.45 $\pm$  0.03 &  9.46 $\pm$  0.03 &  9.43 $\pm$  0.03 &  9.42 $\pm$  0.03 &  9.40 $\pm$  0.04 & \nodata &\nodata &\nodata & 1 &  	III & \nodata\\
 240 & 84.47515 & -2.74462 & 11.92 $\pm$  0.03 & 11.88 $\pm$  0.03 & 11.86 $\pm$  0.03 & 11.82 $\pm$  0.03 & \nodata           & 17.02 $\pm$  0.05 &    q &    0 & \nodata &  	III & 8,9\\
 243 & 84.47665 & -2.65823 &  8.51 $\pm$  0.03 &  8.68 $\pm$  0.03 &  8.55 $\pm$  0.03 &  8.51 $\pm$  0.03 &  8.47 $\pm$  0.03 & 10.91 $\pm$  0.07 &    7 &\nodata & 1 &  	III & 7 \\
 244 & 84.47683 & -2.72711 &  8.99 $\pm$  0.03 &  9.06 $\pm$  0.03 &  8.98 $\pm$  0.03 &  8.96 $\pm$  0.03 &  8.98 $\pm$  0.04 & \nodata &\nodata &\nodata & \nodata &  	III & \nodata\\
 247 & 84.47852 & -2.68587 & 12.04 $\pm$  0.03 & 11.80 $\pm$  0.03 & 11.45 $\pm$  0.03 & 10.82 $\pm$  0.03 &  8.41 $\pm$  0.03 & 18.55 $\pm$  0.06 &    q &    0 & 0 &  	II & 5,9\\
 251 & 84.47960 & -2.46006 & 12.45 $\pm$  0.03 & 12.36 $\pm$  0.03 & 12.30 $\pm$  0.04 & 12.46 $\pm$  0.04 & \nodata           & 18.49 $\pm$  0.10 &    9 &\nodata & \nodata &  	III & 9\\
 254 & 84.48156 & -2.55146 & 13.66 $\pm$  0.03 & 13.35 $\pm$  0.03 & 12.94 $\pm$  0.04 & 12.32 $\pm$  0.04 &  9.70 $\pm$  0.05 & \nodata &\nodata &\nodata & 0 &  	II & 2,3,12\\
 256 & 84.48218 & -2.40933 & 13.79 $\pm$  0.03 & 13.74 $\pm$  0.03 & 13.56 $\pm$  0.06 & 13.77 $\pm$  0.10 & \nodata           & \nodata &\nodata &\nodata & \nodata &  	III & 2,3\\
 260 & 84.48530 & -2.73263 & 11.91 $\pm$  0.03 & 11.96 $\pm$  0.03 & 11.85 $\pm$  0.03 & 11.86 $\pm$  0.03 & \nodata           & 16.84 $\pm$  0.05 &    q &    0 & \nodata &  	III & 9\\
 271 & 84.48935 & -2.64562 & 12.92 $\pm$  0.03 & 12.76 $\pm$  0.03 & 12.58 $\pm$  0.04 & 12.03 $\pm$  0.04 &  9.40 $\pm$  0.04 & \nodata &\nodata &\nodata & 0 &  	II & 1,2,3,6,8,12\\
 275 & 84.49076 & -2.91668 & 11.62 $\pm$  0.03 & 11.62 $\pm$  0.03 & 11.60 $\pm$  0.03 & 11.55 $\pm$  0.03 & \nodata           & 14.86 $\pm$  0.04 &    q &    1 & \nodata &  	III & \nodata\\
 283 & 84.49328 & -2.69059 & 12.06 $\pm$  0.03 & 11.96 $\pm$  0.03 & 11.90 $\pm$  0.03 & 11.96 $\pm$  0.04 & \nodata           & 18.92 $\pm$  0.07 &    q &    0 & 0 &  	III & 6,9,12\\
 293 & 84.49872 & -2.85090 &  9.61 $\pm$  0.03 &  9.70 $\pm$  0.03 &  9.62 $\pm$  0.03 &  9.59 $\pm$  0.03 &  9.67 $\pm$  0.04 & \nodata &\nodata &\nodata & \nodata &  	III & \nodata\\
 297 & 84.50228 & -2.75267 & 11.56 $\pm$  0.03 & 11.47 $\pm$  0.03 & 11.47 $\pm$  0.03 & 11.43 $\pm$  0.03 & \nodata           & 17.75 $\pm$  0.06 &    q &    0 & 1 &  	III & 9,12\\
 299 & 84.50394 & -2.43548 & 11.68 $\pm$  0.03 & 11.76 $\pm$  0.03 & 11.66 $\pm$  0.03 & 11.33 $\pm$  0.03 &  6.76 $\pm$  0.03 & 16.70 $\pm$  0.11 &    q &    1 & \nodata &  	TD$^2$ & 9\\
 300 & 84.50434 & -2.76042 & 10.12 $\pm$  0.03 &  9.70 $\pm$  0.03 &  9.29 $\pm$  0.03 &  8.64 $\pm$  0.03 &  5.87 $\pm$  0.03 & 17.59 $\pm$  0.07 &    q &    1 & \nodata &  	II & 9\\
\enddata
\tablecomments{Table \ref{tab:members} is published in its entirety in the electronic edition of the {\it Astrophysical Journal}. 
A portion is shown here for guidance regarding its form and content.}

\tablenotetext{~}{References in columns 10 and 14: q CIDA survey; 1 \citet{barrado03}; 2 \citet{bejar99}; 3 \citet{bejar01}; 
4 \citet{bejar04}; 5 \citet{burningham05}; 6 \citet{kenyon05}; 7 \citet{kharchenko04}; 8 \citet{oliveira06};
9 \citet{sherry04}; 10 \citet{scholz04}; 11 \citet{zapatero02}; 12 \citet{jeffries06}}
\tablenotetext{~}{Column 11: 1 variable; 0 non-variable}
\tablenotetext{~}{Column 12: 1 X-ray emission; 0 not X-ray emission}
\tablenotetext{~}{Column 13: III=non-excess, II=thick disk, I=class I candidate, EV= evolved disk, 
TD=transition disk candidate, DD=Debris disk candidate}
\tablenotetext{1}{Classified as III in \S \ref{res:slope}}
\tablenotetext{2}{Classified as EV in \S \ref{res:slope}}
\tablenotetext{3}{Classified as II in \S \ref{res:slope}}
\tablenotetext{4}{Classified as I in \S \ref{res:slope}}
\end{deluxetable}

\begin{deluxetable}{lllllllll}
\tabletypesize{\scriptsize}
\tablewidth{0pt}
\tablecaption{Uncertain members of the $\sigma$ Orionis Cluster \label{tab:uncert}}
\tablehead{
\colhead{SO} & \colhead{RA(2000)} & \colhead{DEC(2000)} & \colhead{[3.6]} & \colhead{[4.5]} & \colhead{[5.8]} & \colhead{[8.0]} & \colhead{[24.0]} & \colhead{Disk} \\
\colhead{ID} & \colhead{deg}      & \colhead{deg}       & \colhead{mag}   & \colhead{mag}   & \colhead{mag}   & \colhead{mag}   & \colhead{mag}    & \colhead{type} \\
}
\startdata
  33 & 84.34808 & -2.74109 & 14.49 $\pm$  0.03 & 14.46 $\pm$  0.04 & 14.42 $\pm$  0.07 & 14.44 $\pm$  0.15 & \nodata & III \\
  47 & 84.35836 & -2.34652 & 13.53 $\pm$  0.03 & 13.55 $\pm$  0.03 & 13.47 $\pm$  0.05 & 13.80 $\pm$  0.11 & \nodata & III \\
  57 & 84.36604 & -2.51190 & 13.45 $\pm$  0.03 & 13.45 $\pm$  0.03 & 13.33 $\pm$  0.04 & 13.52 $\pm$  0.08 & \nodata & III \\
  71 & 84.37812 & -2.40533 & 14.45 $\pm$  0.03 & 14.48 $\pm$  0.04 & 14.51 $\pm$  0.11 & 14.21 $\pm$  0.13 & \nodata & III \\
  81 & 84.38234 & -2.40128 & 14.36 $\pm$  0.03 & 14.27 $\pm$  0.04 & 14.19 $\pm$  0.07 & 14.13 $\pm$  0.14 & \nodata & III \\
  87 & 84.38534 & -2.75422 & 13.67 $\pm$  0.03 & 13.70 $\pm$  0.04 & 13.86 $\pm$  0.08 & 13.96 $\pm$  0.15 & \nodata & III \\
 107 & 84.39636 & -2.44933 & 12.10 $\pm$  0.03 & 12.01 $\pm$  0.03 & 11.91 $\pm$  0.03 & 11.94 $\pm$  0.03 & \nodata & III \\
 120 & 84.40674 & -2.42888 & 13.48 $\pm$  0.03 & 13.49 $\pm$  0.03 & 13.35 $\pm$  0.05 & 13.30 $\pm$  0.09 &  9.95 $\pm$ 0.05 & TD$^1$ \\
 219 & 84.46685 & -2.60126 & 13.92 $\pm$  0.03 & 13.81 $\pm$  0.03 & 13.73 $\pm$  0.05 & 13.70 $\pm$  0.07 & \nodata & III$^6$ \\
 127 & 84.40806 & -2.80576 & 12.78 $\pm$  0.03 & 12.82 $\pm$  0.03 & 12.80 $\pm$  0.04 & 12.73 $\pm$  0.04 & \nodata & III \\
 134 & 84.41515 & -2.30742 & 14.14 $\pm$  0.03 & 13.93 $\pm$  0.03 & 13.87 $\pm$  0.06 & 13.74 $\pm$  0.08 & \nodata & III \\
 154 & 84.43077 & -2.44282 & 14.48 $\pm$  0.03 & 14.45 $\pm$  0.04 & 14.35 $\pm$  0.08 & 14.15 $\pm$  0.09 & \nodata & III \\
 174 & 84.44174 & -2.45832 & 13.71 $\pm$  0.03 & 13.63 $\pm$  0.03 & 13.61 $\pm$  0.05 & 13.53 $\pm$  0.08 & \nodata & III \\
 188 & 84.45503 & -2.36186 & 14.14 $\pm$  0.03 & 14.17 $\pm$  0.04 & 13.94 $\pm$  0.06 & 14.06 $\pm$  0.13 & \nodata & III \\
 207 & 84.46257 & -2.32070 & 14.13 $\pm$  0.03 & 14.18 $\pm$  0.03 & 14.14 $\pm$  0.07 & 14.20 $\pm$  0.11 & \nodata & III \\
 221 & 84.46809 & -2.44730 & 14.24 $\pm$  0.03 & 14.22 $\pm$  0.03 & 14.29 $\pm$  0.07 & 14.10 $\pm$  0.12 & \nodata & III \\
 236 & 84.47451 & -2.48824 & 14.48 $\pm$  0.03 & 14.46 $\pm$  0.04 & 14.07 $\pm$  0.07 & 14.13 $\pm$  0.13 & \nodata & III \\
\enddata
\tablecomments{Table \ref{tab:uncert} is published in its entirety in the electronic edition of 
the {\it Astrophysical Journal}. A portion is shown here for guidance
regarding its form and content.}
\tablenotetext{~}{Column 9: III=non-excess, II=thick disk, I=class I candidate, EV= evolved disk, TD=transition disk candidate, 
DD=Debris disk candidate}
\tablenotetext{1}{Classified as III in \S \ref{res:slope}}
\tablenotetext{2}{Classified as EV in \S \ref{res:slope}}
\tablenotetext{3}{Classified as II in \S \ref{res:slope}}
\tablenotetext{4}{Classified as I in \S \ref{res:slope}}
\tablenotetext{5}{Possible giant in the J-H versus H-K color-color diagram}
\tablenotetext{6}{Radial velocity consistent for stars in the older population \citep{jeffries06}}
\end{deluxetable}

\clearpage

\begin{deluxetable}{ccccc}
\tabletypesize{\scriptsize}
\tablewidth{0pt}
\tablecaption{Median SEDs and Quartiles of optically thick disk stars \label{tab:sed_ctts}}
\tablehead{
\colhead{Wavelength} & \colhead{Median} & \colhead{Upper} & \colhead{Lower} & \colhead{stars}\\
\colhead{\micron } & \colhead{log $\frac{\lambda F_\lambda}{1.24 F_J}$} & \colhead{} &  \colhead{} &  \colhead{used}
}
\startdata
0.44 & -1.506 & -1.254 & -1.654 & 22\\
0.55 & -1.002 & -0.842 & -1.274 & 51\\
0.64 & -0.738 & -0.588 & -0.928 & 60\\
0.79 & -0.308 & -0.220 & -0.376 & 52\\
1.24 & 0.0 & 0.0 & 0.0 & 64\\
1.66 & -0.037 & -0.005 & -0.065 & 64\\
2.16 & -0.209 & -0.125 & -0.245 & 64\\
3.6 & -0.563 & -0.400 & -0.645 & 64\\
4.5 & -0.733 & -0.580 & -0.836 & 64\\
5.8 & -0.910 & -0.748 & -0.978 & 64\\
8.0 & -0.962 & -0.804 & -1.076 & 64\\
24 & -1.238 & -1.029 & -1.430 & 58 \\
\enddata
\end{deluxetable}

\begin{deluxetable}{ccccc}
\tabletypesize{\scriptsize}
\tablewidth{0pt}
\tablecaption{Median SEDs and Quartiles of non-excess stars \label{tab:sed_wtts}}
\tablehead{
\colhead{Wavelength} & \colhead{Median} & \colhead{Upper} & \colhead{Lower} & \colhead{stars}\\
\colhead{\micron } & \colhead{log $\frac{\lambda F_\lambda}{1.24 F_J}$} & \colhead{} &  \colhead{} &  \colhead{used}
}
\startdata
0.44 & -1.360 & -1.206 & -1.550 & 60 \\
0.55 & -0.863 & -0.610 & -1.166 & 116 \\
0.64 & -0.597 & -0.400 & -0.870 & 128 \\
0.79 & -0.204 & -0.144 & -0.324 & 97 \\
1.235 & 0 & 0 & 0 & 137 \\
1.662 & -0.057 & -0.037 & -0.073 & 137 \\
2.159 & -0.257 & -0.241 & -0.273 & 137 \\
3.6 & -0.748 & -0.733 & -0.777 & 137 \\
4.5 & -1.038 & -1.007 & -1.083 & 137 \\
5.8 & -1.306 & -1.278 & -1.348 & 137 \\
8.0 & -1.693 & -1.660 & -1.723 & 137 \\
24 & -3.259 & -3.198 & -3.431 & 42$^1$ \\
\enddata
\tablenotetext{1}{Including stars in the mass range of IMTTS and HAeBe}
\end{deluxetable}

\begin{deluxetable}{ccc}
\tabletypesize{\scriptsize}
\tablewidth{0pt}
\tablecaption{Disk fractions in the $\sigma$ Orionis cluster \label{tab:frac}}
\tablehead{
\colhead{Range} & \colhead{Thick Disks} & \colhead{Thick+Evolved Disks} \\
\colhead{ } & \colhead{ \% } & \colhead{\%} 
}
\startdata
HAeBe      & 3.7$\pm$3.7  & 14.8$\pm$7.4 \\
IMTTS      & 19.6$\pm$5.9  & 26.8$\pm$6.9 \\
TTS        & 31.1$\pm$3.8  & 36.3$\pm$4.1 \\
BDwarfs     & 29.7$\pm$9.0  & 33.3$\pm$9.7 \\
Total      & 26.6$\pm$2.8  & 33.9$\pm$3.1 \\
\enddata
\end{deluxetable}

\clearpage

\begin{figure}
\epsscale{1.0}
\plotone{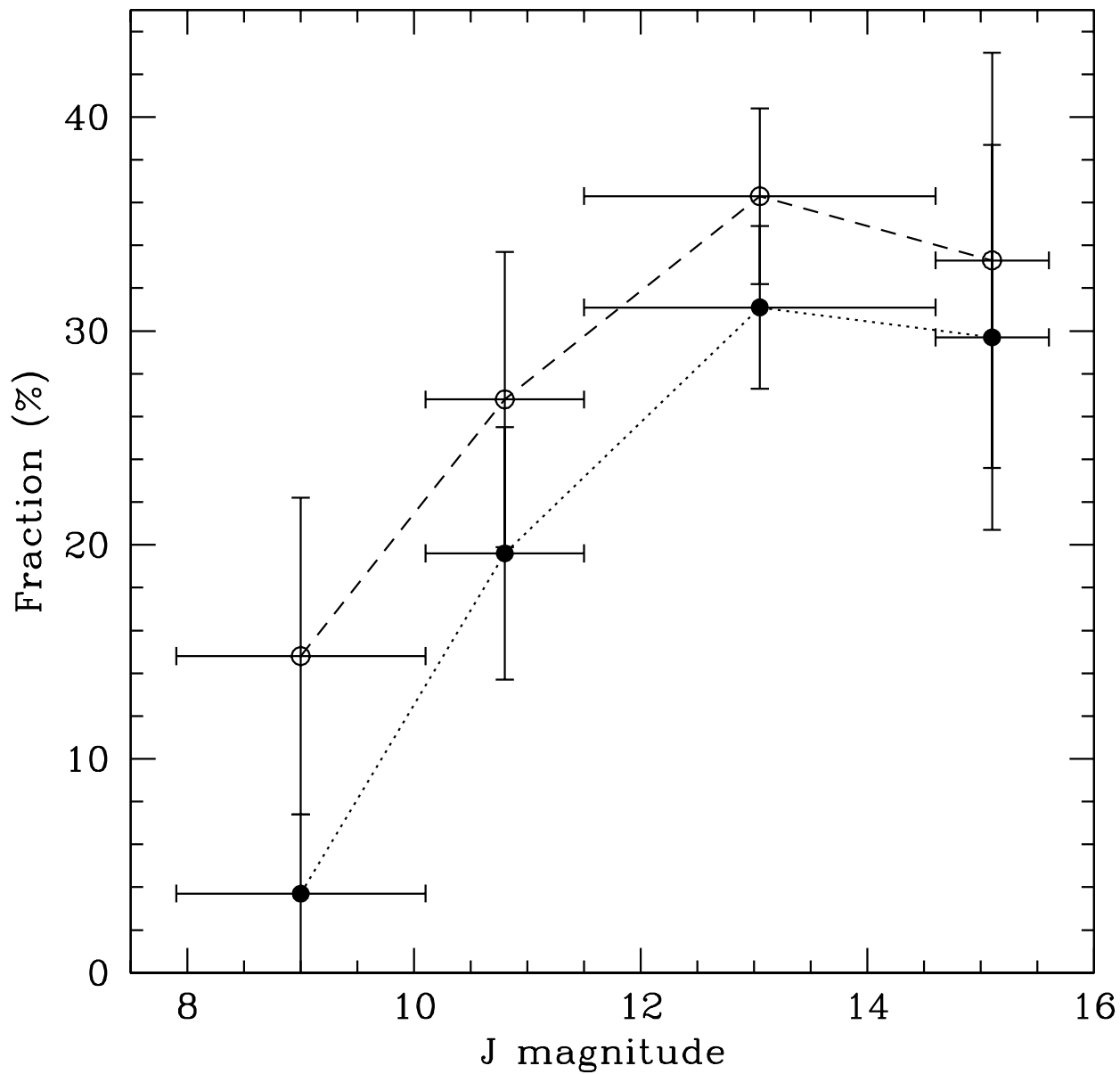}
\caption{Disk fraction versus J magnitude. 
Each point represents 
a different mass range (see Figure \ref{fig:alphaJ}): from 
left-to-right, HAeBe, IMTTS, TTS and Brown Dwarfs. 
Solid circles + dotted line indicate thick disks fraction. 
Open circles + dashed line are the fraction of disk-bearing stars. 
The lowest disk fraction is observed in the HAeBe mass range, while the highest 
frequency of disks is observed in the TTS mass range.}
\label{fig:fracJ} 
\end{figure}

\begin{figure}
\epsscale{1.0}
\plotone{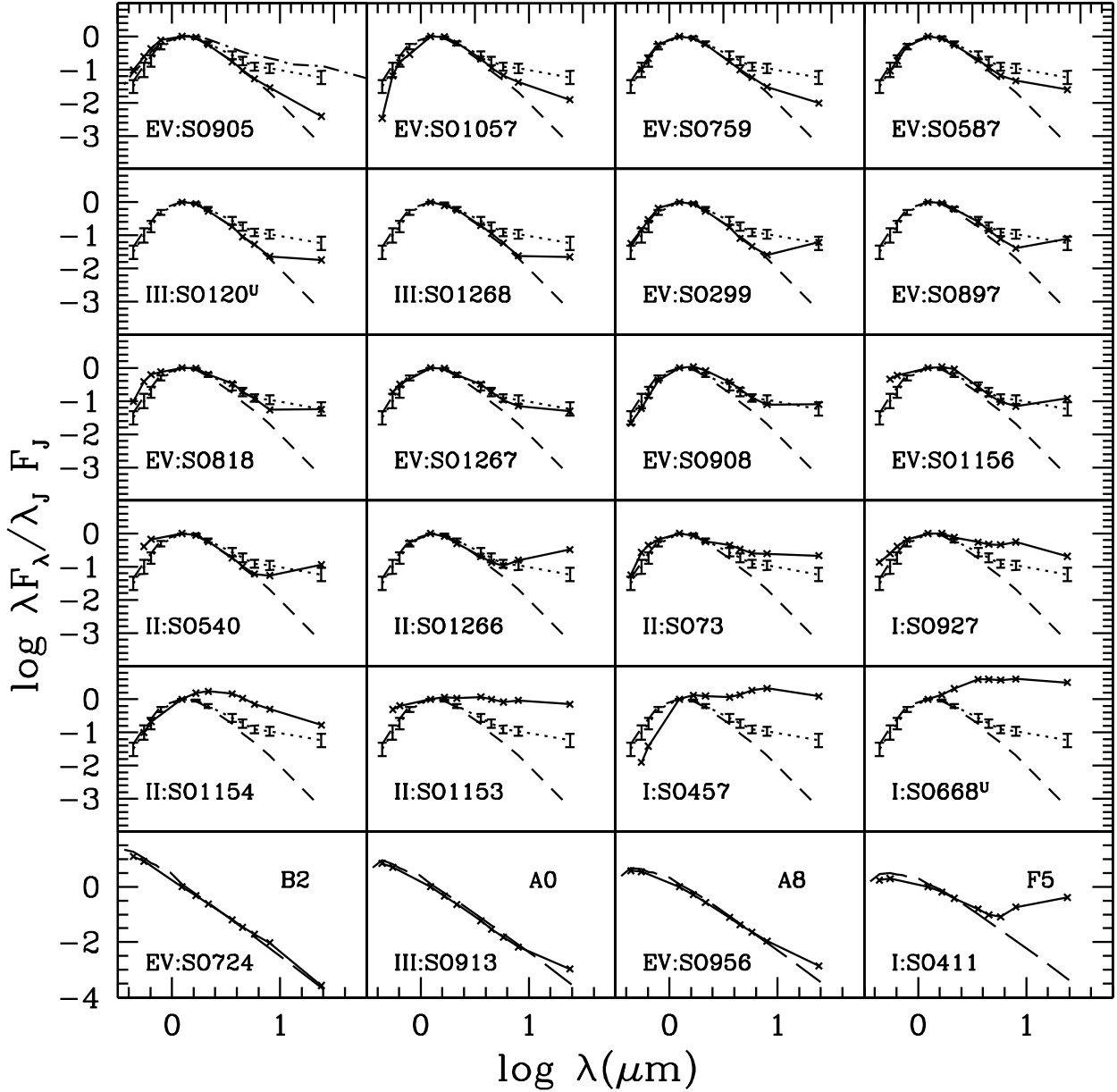}
\caption{Spectral energy distributions for  selected stars, illustrating 
the diversity of disks found in the $\sigma$ Orionis Cluster. The first 5 rows show
stars in the mass range of TTS. 
Dotted and dashed lines represent the median SED of optically thick disks and non-excess stars 
of the $\sigma$ Orionis cluster, respectively. Error bars denote the quartiles 
of the optically thick disk median SED. Dot-dashed line, shown in the first panel, 
represents  the median SED of optically thick disks in Taurus. The last row shows 
stars in the mass range of HAeBe; these are compared with the photosphere of a star
with a similar spectral type \citep[long-dashed line; ][]{kh95}. Each panel is 
labeled with the internal running identification number and the classification from
\S \ref{res:slope}. }
\label{fig:sed}
\end{figure}

\begin{figure}
\epsscale{1.0}
\plotone{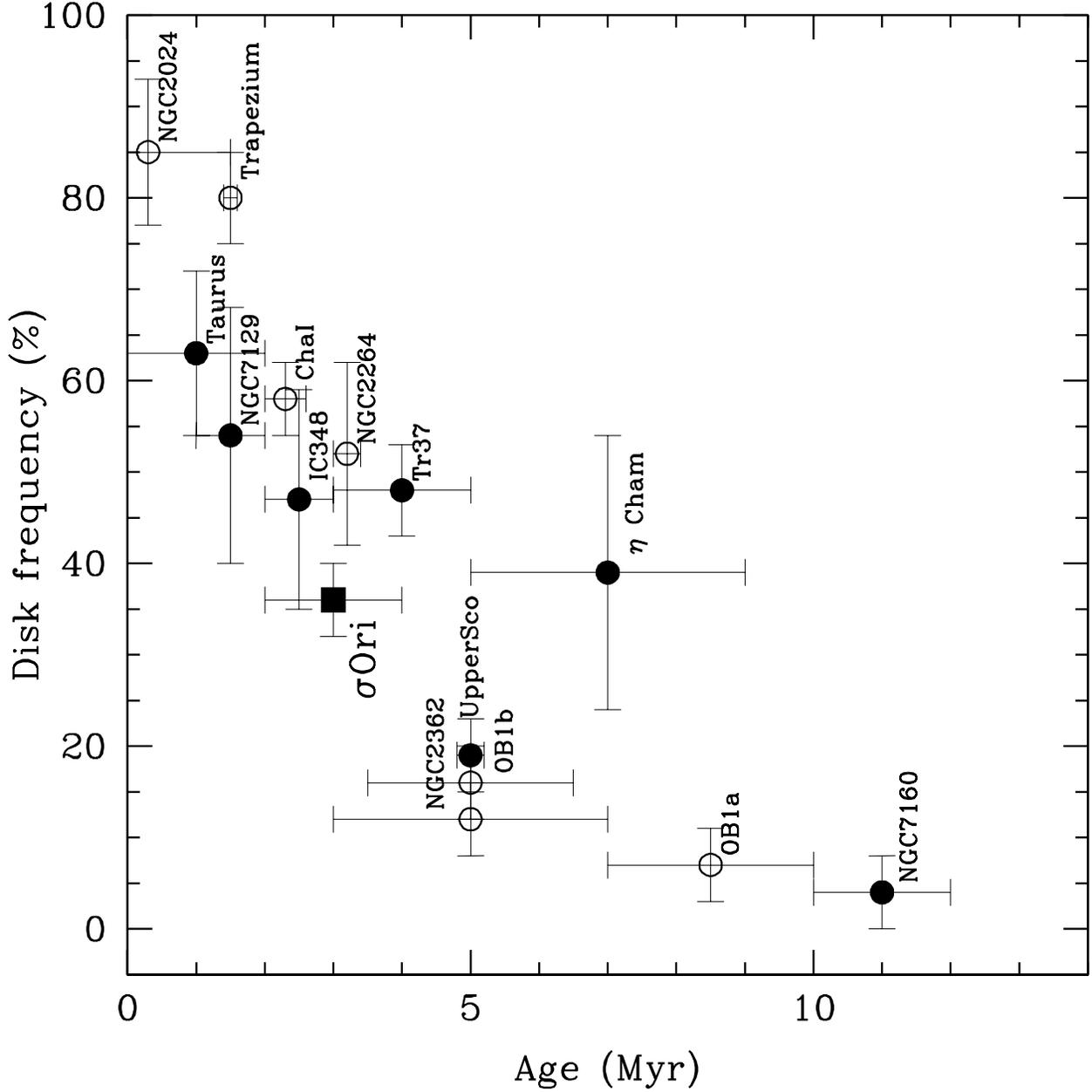}
\caption{ Fraction of stars with near-infrared disk emission as a function 
of the age of the stellar group.
Open circles represent the disk frequency for stars in the TTS mass range, derived using JHKL observations: 
NGC2024, Trapezium, NGC2264 and NGC2362 from \citet{haisch01},  Chamaleon I from \citet{gomez01},
and OB1a and OB1b sub-associations from \citet{hernandez05}. Solid symbols represent the disk frequency
calculated for stars in the TTS mass range using IRAC data: Taurus \citep{hartmann05c}, 
NGC7129 \citep{gutermuth04}, IC348 \citep{lada06}, Tr 37 and NGC7160 \citep{aurora06}, 
Upper Scorpius \citep{carpenter06},
$\eta$ Chameleontis \citep{megeath05a} 
and $\sigma$ Orionis cluster (this work) }
\label{fig:disk_evol}
\end{figure}

\clearpage

\begin{figure}
\epsscale{1.0}
\plotone{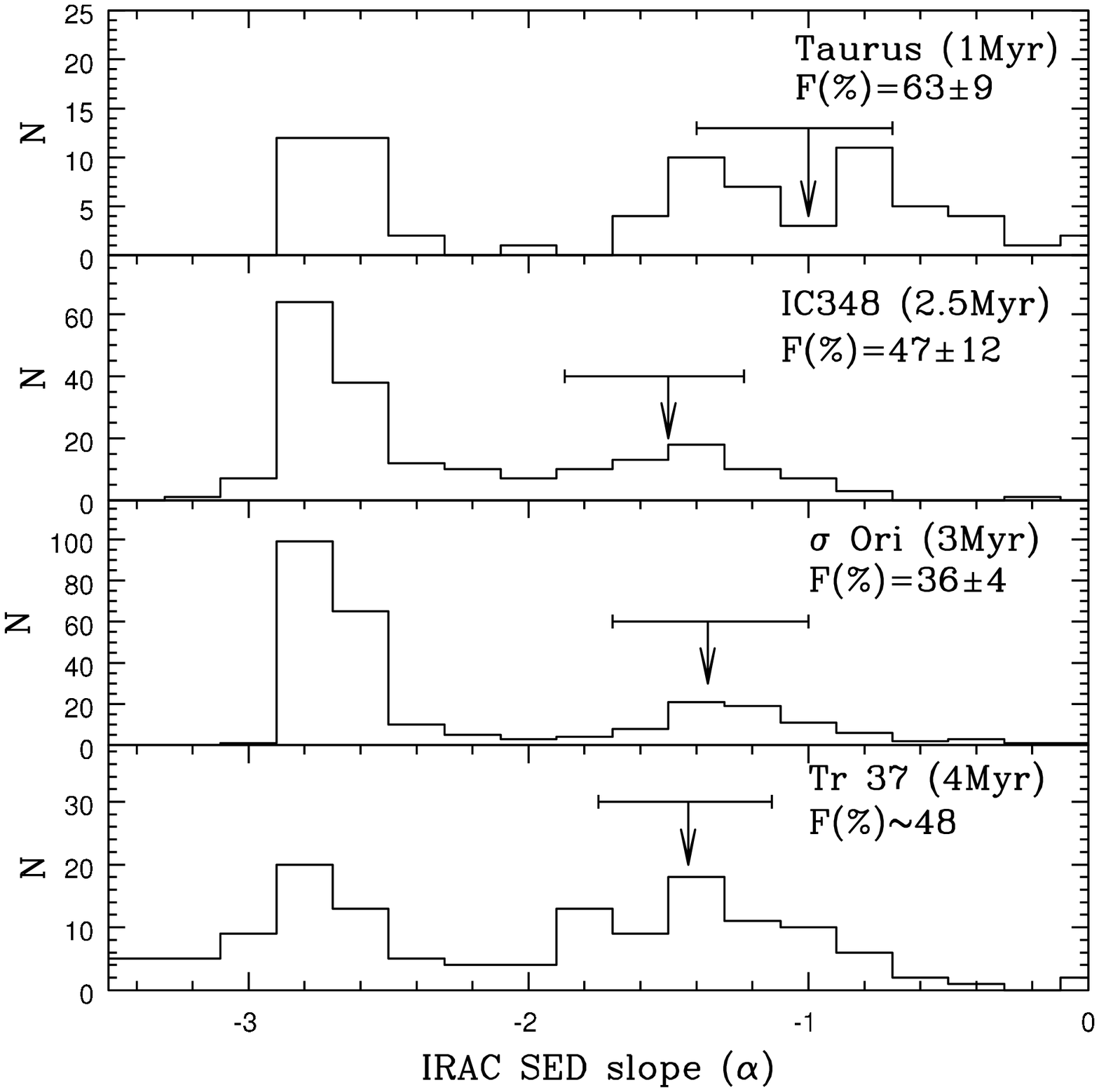}
\caption{IRAC SED slope distribution for young stellar groups with differing ages.
The dotted line defines the limit between thick disks and evolved disks. Fractions 
of disk-bearing stars and stellar ages are been included in each panel:
Taurus \citep{hartmann05c}, IC348 \citep{lada06}, $\sigma$ Orionis (this work) and
Tr 37 \citep{aurora06}. The median of $\alpha$  slope and its quartiles for 
disk-bearing stars in each stellar group are indicated with arrows and error bars, respectively.}
\label{fig:slopehist}
\end{figure}

\begin{figure}
\epsscale{0.8}
\plotone{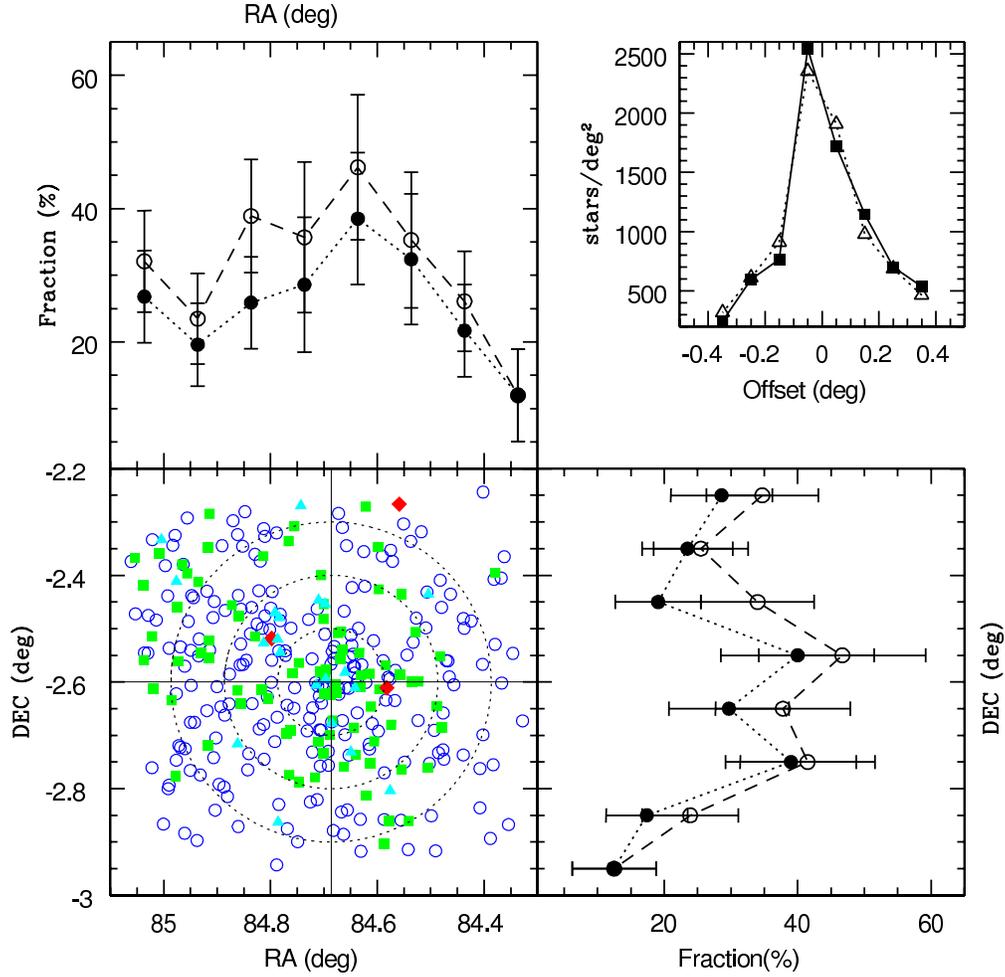}
\caption{Diagrams illustrating the space distribution of members, disk fractions 
and the density of the cluster. The lower-left panel shows the location 
of members in the $\sigma$ Orionis cluster. Symbols are similar to Figure \ref{fig:irac_cc}.
Dotted circles define radial distances (at 0.1, 0.2 and 0.3 degree) from the center 
of the cluster and solid lines define the semi-circles East-West (vertical) and 
North-South (horizontal) used to calculate the fraction of disks in bins at several radial 
distance from the star $\sigma$ Ori. Upper left panel shows the fraction of thick disk stars
(filled circles + dotted line) and disk-bearing stars (open circles + dashed line) 
in radial distance bins defined by the semicircles East-West. Similarly, lower-right panel
shows the fraction of disks in radial distance bins defined by the semicircles North-South.
Upper-right panel shows the density of members in radial distance bins defined by 
the semicircles East-West (filled squares+solid line) and by the semicircles 
North-South (open triangles+dotted line) [See electronic edition of the journal for a color 
version of this Figure] }
\label{fig:frac1}
\end{figure}

\begin{figure}
\epsscale{1.0}
\plotone{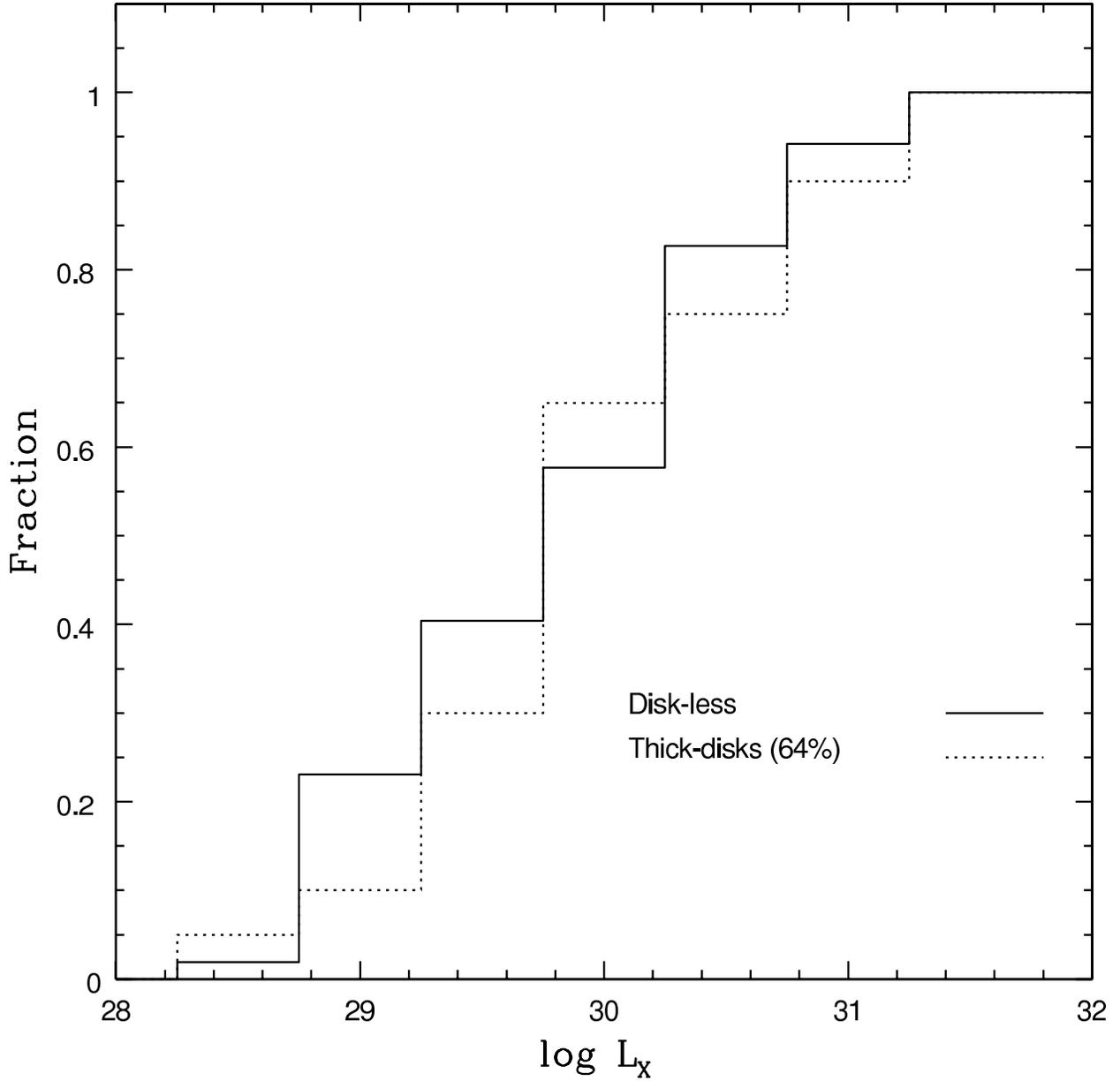}
\caption{Cumulative distribution of non-excess stars and 
optically thick disk stars as a function of X-ray luminosity. 
A Kolmogorov-Smirnov (KS) test shows a significance level of 64\% 
when comparing both populations. This suggests that the X-ray emission 
has a similar origin in non-excess stars (like WTTS) and stars with optically thick disks (like CTTS)}. 
\label{fig:alphaX}
\end{figure}

\end{document}